\documentclass[letterpaper]{article}
\usepackage{aaai}
\usepackage{times}
\usepackage{helvet}
\usepackage{courier}
\usepackage{graphicx}
\usepackage{caption}
\usepackage{subcaption}
\usepackage{amsmath,amssymb}
\usepackage{url}
\usepackage{MnSymbol}

\DeclareFixedFont{\auacc}{OT1}{phv}{m}{n}{12}
\DeclareFixedFont{\afacc}{OT1}{times}{m}{n}{10}
\makeatletter
\newcommand*{\rom}[1]{\expandafter\@slowromancap\romannumeral #1@}
\makeatother

\author{
 Desislava Hristova$^*$, Anastasios Noulas$^*$, Chlo\"{e} Brown$^*$, Mirco Musolesi$^{**}$, Cecilia Mascolo$^*$\\
 $^*$Computer Laboratory, University of Cambridge\\
 $^{**}$School of Computer Science, University of Birmingham\\
 \texttt{dh475,an346,clb76,cm542 @ cam.ac.uk; m.musolesi @ cs.bham.ac.uk}
}

\date{}

\begin{document}

\title{A Multilayer Approach to Multiplexity and Link Prediction in \\Online Geo-Social Networks}
\maketitle

\begin{abstract}
Online social systems are multiplex in nature as multiple links may exist between the same two users across different social networks. In this work, we introduce a framework for studying links and interactions between users beyond the individual social network. Exploring the cross-section of two popular online platforms - Twitter and location-based social network Foursquare - we represent the two together as a composite \emph{multilayer online social network}. Through this paradigm we study the interactions of pairs of users differentiating between those with links on one or both networks. We find that users  with multiplex links, who are connected on both networks, interact more and have greater neighbourhood overlap on both platforms, in comparison with pairs who are connected on just one of the social networks. In particular,  the most frequented locations of users are considerably closer, and similarity is considerably greater among multiplex links. We present a number of structural and interaction features, such as the multilayer Adamic/Adar coefficient, which are based on the extension of the concept of the node neighbourhood beyond the single network. Our evaluation, which aims to shed light on the implications of multiplexity for the link generation process, shows that multilayer features, constructed from properties across social networks, perform better than their single network counterparts in predicting links across networks. 
We propose that combining information from multiple networks in a multilayer configuration can provide new insights into user interactions on online social networks, and can significantly improve link prediction overall with valuable applications to social bootstrapping and friend recommendations.
 \end{abstract}

\section{Introduction}

Online social media has become an ecosystem of overlapping and complementary social networking services, inherently multiplex in nature, as multiple links may exist between the same pair of users~\cite{kivela2013}. Multiplexity is a well studied property in the social sciences~\cite{haythornthwaite1998} and it has been explored in social networks from Renaissance Florence~\cite{Padgett06organizationalinvention} to the Internet age~\cite{hay2005}. Despite the broad contextual differences, multi-relational ties are consistently found to exhibit greater intensity of interactions across different communication channels, and therefore a stronger bond~\cite{haythornthwaite1998,me2014}. Nevertheless,  \emph{there is a lack of research about online social networks and their value from a multiplex perspective}.

Recently, empirical models of multilayer networks have emerged to address the multi-relational nature of social networks~\cite{kivela2013,Szell19072010}. In such models, interactions are considered as layers in a systemic view of the social network. We adopt such a model in our analysis, where we shift the concept of a link and neighbourhood to encompass more than one network. This allows us to study interactions and structural properties across online social networks (OSNs), addressing the need for further understanding of their complimentary and overlapping nature, and multiplexity online. Although there have been some recent comparative studies of multiple online social networks~\cite{ICWSM148059,lee2014}, and their intersection~\cite{Szell19072010}, the applications of multiplex network properties to OSNs is yet to be substantially addressed.  

In this work, we explore intersecting networks, multiplex ties, and their application to link prediction across OSNs. Link prediction systems are key components of social networking services due to their practical applicability to friend recommendations and social network bootstrapping, as well as to understanding the link generation process. Link prediction is a well-studied problem, explored in the context of both OSNs and location-based social networks (LBSNs)~\cite{liben2007link,menon2011link,crandall2010,scellato2011exploiting}.  However, only very few link prediction works tackle multiple networks at a time~\cite{lee2014,Tang2012}, while \emph{most link prediction systems only employ features internal to the network under prediction}, without considering additional link information from other OSNs.

Our main contributions can be summarised as follows:

\begin{itemize}
\item We generalise the notion of a \emph{multilayer online social network}, and extend definitions of neighbourhood to span multiple networks,  adapting measures of overlap such as the Adamic/Adar coefficient in social networks to the multilayer context. \\ 


\item  We find that \emph{pairs with links on both Twitter and Foursquare exhibit significantly higher interaction on both social networks} in terms of number of mentions and colocation within the same venues, as well as a lower distance and higher number of common hashtags in their tweets. 

\item  A significantly \emph{higher overlap can be observed between the neighbourhoods of nodes with links on both networks}, in particular with relation to the Adamic/Adar measure of neighbourhood overlap, which is significantly more expressed in the multilayer neighbourhood. 

\item In our evaluation, \emph{we predict Twitter links from Foursquare features and vice versa}, and we achieve this with AUC scores up to 0.86 on the different datasets. In predicting links which span both networks, we achieve the highest AUC score of 0.88 from our multilayer features set, \emph{proving the multilayer construct a useful tool for social bootstrapping and friend recommendations}. 
\end{itemize}

The remainder of this work details these contributions, and summarises related work, concluding with a discussion of the implications, limitations, and applications of the proposed framework.

\section{Related Work}

Our work identifies with three main areas:  multi-relational social networks, media multiplexity, and link prediction in online social networks. We summarise the state of the art in these areas in the following sections.

\subsection{Multilayer Social Networks}
Multi-relational or multilayer networks have been explored in the context of a wide range of systems from global air transportation~\cite{cardillo2013} to massive online multiplayer games~\cite{Szell19072010}. A comprehensive review of multilayer network models can be found in~\cite{kivela2013}. In the context of social networks, it is generally accepted that the more information we can obtain about the relationship between people, the more insight we can gain. A recent large-scale study on the subject has demonstrated the need for multi-channel data when comprehensively studying social networks~\cite{lehman2014}. Despite the observable multilayer nature of the composite OSNs of users~\cite{kivela2013,kazienko2010,brodka2012},  most research efforts have been focused on theoretical modelling~\cite{kivela2013}, with little to no empirical work exploiting data-driven applications in the domain of multilayer OSNs, especially with respect to how location-based and social interactions are coupled in the online social space. We attempt to fill these gaps in the present work by presenting a generalisable online multilayer framework applied to classic problems such as link prediction in OSNs. Our framework is strongly motivated by the theory of media multiplexity, which we review next.

\subsection{Media Multiplexity}
Media multiplexity~\cite{hay2005} is the principle that tie strength is observed to be greater when the number of media channels used to communicate between two people is greater (higher multiplexity). In~\cite{haythornthwaite1998} the authors studied the effects of media use on relationships in an academic organisation and found that those pairs of participants who utilised more types of media (including email and videoconferencing) interacted more frequently and therefore had a closer relationship, such as friendship. More recently, multiplexity has been studied in light of multilayer communication networks, where the intersection of the layers was found to indicate a strong tie, while single-layer links were found to denote a weaker relationship~\cite{me2014}. 
The strength of social ties is an important consideration in friend recommendations and link prediction~\cite{Gilbert2009}, and we employ the previously understudied multiplex properties of OSNs to such ends in this work. 

\subsection{Link Prediction}
The problem of link prediction was first introduced in the seminal work of Kleinberg et al.~\cite{liben2007link} and since then, has been applied in various network domains. For instance, in~\cite{scellato2011exploiting} the authors exploit place features in location-based services to recommend friendships, and in~\cite{backstrom2011supervised} a new model based on supervised random walks is proposed to predict new links in Facebook. 
Most of these works  build on features that are endogenous to the system that hosts the social network of users. In our evaluation, however, we train and test on heterogeneous networks. In a similar spirit,  the authors in 
~\cite{Sadilek2012} show how using both location and social information from the same network significantly improves link prediction. Our approach differs in that it frames the link prediction task in the context of multilayer networks and empirically shows the relationship between two different systems - Foursquare and Twitter - by mining features from both. 
Before presenting our framework and analysis, we will next state the research questions we are interested in answering through this work. 

\section{Research Questions}

In light of the related work presented above, our goal is to mend the gap between multilayer network models, media multiplexity properties, and link prediction systems. More specifically, we address the following research questions in this work: \\

\textbf{RQ1:} \emph{How do structural properties such as degree extend into the multilayer neighbourhood?} We propose a multilayer version of the network neighbourhood, which extends it to multiple networks (layers) and observe how such structural properties are manifested across Twitter and Foursquare.\\
\\
\textbf{RQ2:} \emph{What are the structural and behavioural differences between single network and multiplex links?} In order to understand the value of multiplex links (users connected on more than one network), we observe how they compare to single network links in terms of neighbourhood overlap, Twitter interaction, similarity and mobility in Foursquare.\\
\\
\textbf{RQ3:} \emph{Can we use information about links from one layer to predict links on the other?} Many online social systems suffer from a lack of initial user adoption. Although many social networks nowadays incorporate the option of importing contacts from another pre-existing network and copying links, this method does not offer a ranking of users by relevance targeted towards the specific platform.\\
\\
 \textbf{RQ4:} \emph{Can we predict links which exist on more than one network (i.e., multiplex links)?} Media multiplexity is a valuable source of tie strength information, and has further structural implications, which are of interest to OSN services and link prediction systems. We would like to explore the potential of identifying such links for building more successful online communities.\\
\\
We will next present our multilayer framework for OSNs, and study user behaviour and properties across Twitter and Foursquare, extending our analysis to multiplex links in comparison with single-layer links. We finally integrate this into a link prediction system for OSNs, where we evaluate the utility of the metrics and features described in this work in hope to answer the above posed questions. 

\section{Multi-relational Framework}

The network of human interactions is usually represented by a graph $G$ where the nodes in set $V$ represent people and the edges $E$ represent interactions. While this representation has been immensely helpful for the uncovering of many social phenomena, it is focused on a single-layer abstraction of human relations. In this section, we describe a model, which represents the multiplexity of OSNs by supporting multiple friendship and interaction links.

\subsection{Multilayer Online Social Network}

\begin{figure}[t!]
\centering
 \begin{subfigure}[b]{0.24\textwidth}
                \includegraphics[width=\textwidth]{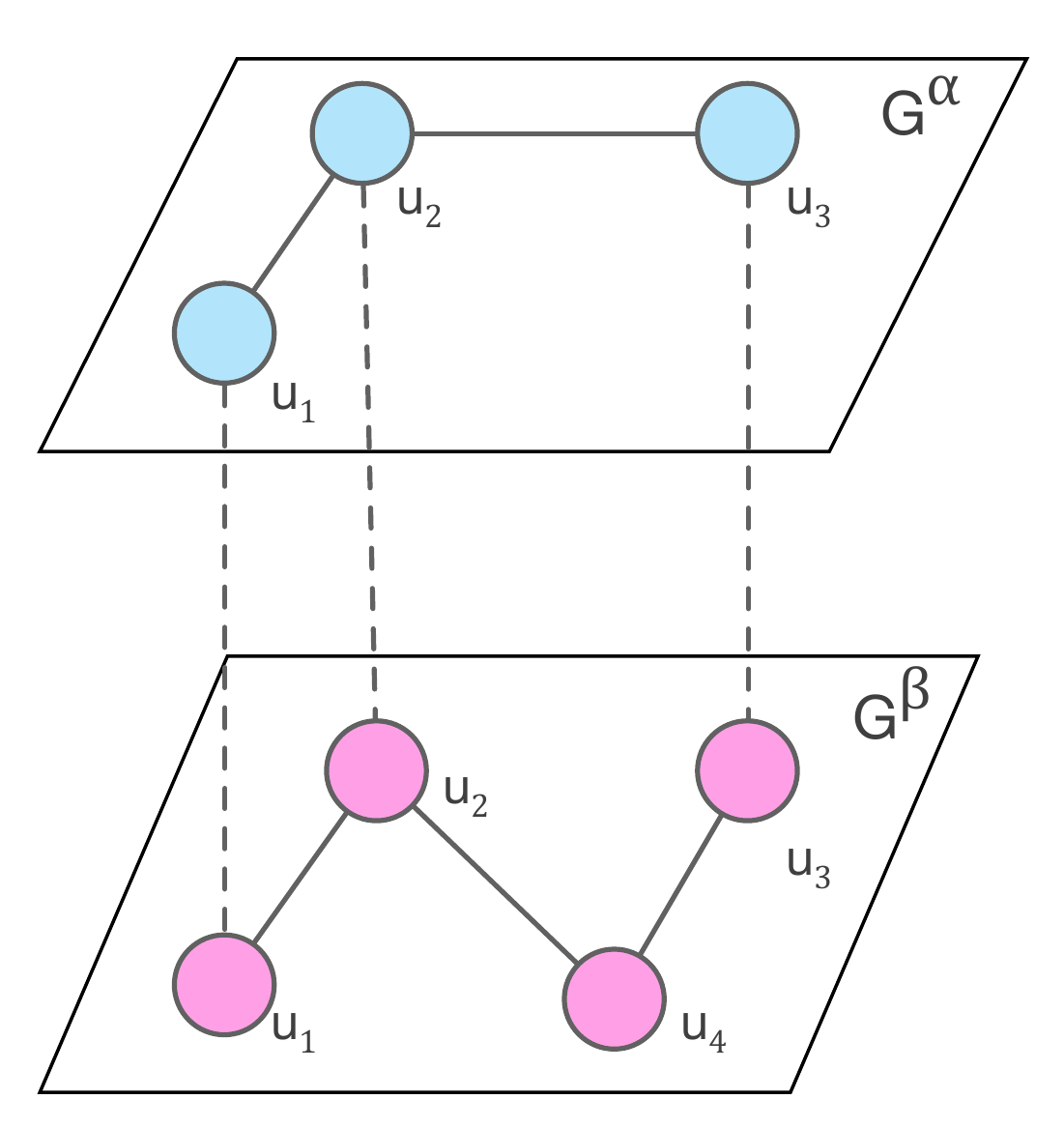}
                \caption{Multilayer Social Network}
                \label{fig:multi}
        \end{subfigure}%
         \begin{subfigure}[b]{0.26\textwidth}
                \includegraphics[width=\textwidth]{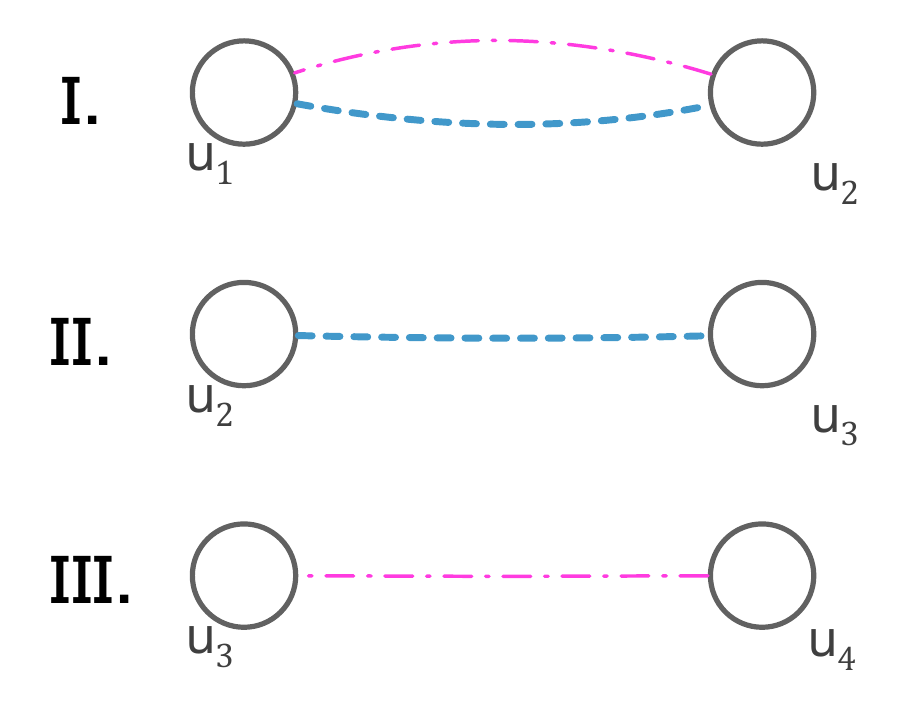}
                \caption{Link types}
                \label{fig:links}
        \end{subfigure}%
        \caption{Multilayer model of OSNs with \rom{1}. Multiplex link; \rom{2}. Single-layer link on $G^\alpha$; and \rom{3}. Single-layer link on $G^\beta$.}
          \label{fig:mdat}
\end{figure}

We represent the parallel interactions between nodes across OSNs as a \emph{multilayer network} $\cal{M}$, an ensemble of $M$ graphs, each corresponding to an OSN. We indicate the $\alpha$-th layer of the multilayer  as $G^\alpha(V^\alpha, E^\alpha)$, where $V^{\alpha}$ and $E^{\alpha}$ are the sets of vertices and edges of the graph $G^{\alpha}$. 
We can then denote the sequences of graphs composing the $M$-layer multilayer graph as ${\cal{M}} = \{G^1,...,G^\alpha,...,G^M\}$. The graphs are brought together as a multilayer system by the common members across layers as 
illustrated in Figure~\ref{fig:mdat}.

Multilayer social networks are a natural representation of media multiplexity, as each layer can depict an OSN.
Figure~\ref{fig:multi} illustrates the case at hand, where there are two OSN platforms represented by $G^\alpha$ and $G^\beta$. Members need not be present at all layers and the multilayer network is not limited to two layers. While each platform can be explored separately as a network in its own right, this does not capture the dimensionality of online social life,  which spans across multiple OSNs. 

Figure~\ref{fig:links} illustrates three link types as observed in Figure~\ref{fig:multi} for the case of a two layer network. Firstly, we define a \textit{multiplex link} between two nodes $i$ and $j$ as a link that exists between them \emph{at least in two layers} $\alpha, \beta \in \cal{M}$. Second, we say that a \textit{single-layer link} between two nodes $i$ and $j$ exists if the link appears \emph{only in one layer} in the multilayer social network. In systems with more layers, multiplexity can take on a value depending on how many layers the link is present on. 
In the case at hand, given layer $\alpha$ and layer $\beta$, we denote the set of all links present in the multilayer network as $E^{\alpha \cup \beta}$, which yields the global connectivity.

We also define the set of multiplex links as $E^{\alpha \cap \beta}$ and the set of all single-layer links on layer $\alpha$ only as $E^{\alpha \backslash \beta}$. These multilayer edge sets can be further extended to the $M$ layer network by considering more layers $\{1, \dots, M\}$ as part of the intersection or union of graphs.
The presence of multiplex and single-layer links in the above edge sets defines the multilayer neighbourhood of nodes in the network, as expanded upon next. 

\subsection{The Multilayer Neighbourhood}

Following our definition of a multilayer online social network, we can redefine the ego network  of a node as the \emph{multilayer neighbourhood}. While the simple node neighbourhood is the collection of nodes one hop away from the ego, the multilayer global neghbourhood (denoted by $GN$) of a node $i$  can be derived by the total number of unique neighbours across layers:

\begin{equation}
\Gamma_{GNi} = \{j \in V^{\cal{M}} : e_{i,j} \in E^{\alpha \cup \beta}\}
\end{equation}

and their global multilayer degree as:

\begin{equation}
k_{GNi} = | \Gamma_{GNi}|
\end{equation}

which provides insight into the entire connectivity of nodes across layers, and can therefore be interpreted as a global measure of the immediate degree of a node. We can similarly define the core neighbourhood (denoted by $CN$) of a node $i$ across layers of the multilayer network as: 

\begin{equation}
\Gamma_{CNi} = \{j \in V^{\cal{M}} : e_{i,j} \in E^{\alpha \cap \beta}\}
\end{equation}

and their core multilayer degree as:

\begin{equation}
k_{CNi} = | \Gamma_{CNi}|
\end{equation}

where we only consider neighbours which exist across all layers.
This simple formulation allows for powerful extensions of existing metrics of local neighbourhood similarity. We can define the overlap (Jaccard similarity) of two users $i$ and $j$'s global neighbourhoods as:

\begin{equation}
sim_{GNij} = \frac{|\Gamma_{GNi} \bigcap \Gamma_{GNj}|}{|\Gamma_{GNi} \bigcup \Gamma_{GNj}|}
\end{equation}

where the number of common friends is divided by the number of total friends of $i$ and $j$. The same can be done for the core degree of two users. The Jaccard coefficient, often used in information retrieval, has also been widely used in link prediction~\cite{liben2007link}.

We can further extend our definition of the multilayer neighbourhood to the Adamic/Adar coefficient for link likelihood~\cite{adamic2001}, which considers the overlap of two neighbourhoods based on the popularity of common friends (originally through web pages) in a single-layer network as:

\begin{equation}
aa\_sim_{GNij} = \sum_{z \in \Gamma_{GNi} \bigcap \Gamma_{GNj}} 
	\frac{1}{log(|\Gamma_{GNz}|)}
\end{equation}

where it is applied to the global common neighbours between two nodes but can be equally applied to their core neighbourhoods. This metric has shown to be successful in the link prediction in its original single-layer form in both social networks and location-based networks~\cite{liben2007link,scellato2011exploiting}. In the present work, we aim to show its applicability to the multilayer space in predicting online social links across and between Twitter and Foursquare. We will next describe the specific datasets, which we apply this framework to.

\section{Dataset}

Twitter and Foursquare are two of the most popular social networks, both with respect to research efforts and user base. They have  distinct broadcasting functionalities - microblogging and check-ins. While Twitter can reveal a lot about user interests, Foursquare check-ins provide a proxy for human mobility. 
In Foursquare users check-in to venues  that they visit through their location enabled devices, and share their visit or opinion of a place with their friends. Foursquare is two years younger than Twitter and its broadcasting functionality is exclusively for mobile users (50M to date\footnote{\url{https://foursquare.com/about}}), while 80\% of Twitter's 284M users are active on mobile\footnote{\url{https://about.twitter.com/company}}. Twitter generally allows anyone to ``follow" and be ``followed", where followers and followed do not necessarily know one another. On the other hand, Foursquare supports undirected links, referred to as ``friendship". A similar undirected relationship can be constructed from Twitter, where a link can be considered between two users if they both follow each other reciprocally~\cite{Kwak2010}. Since we are interested in ultimately in predicting friendship, we consider only reciprocal Twitter links throughout this work.

\begin{figure}[t!]
\centering
                \includegraphics[scale=0.17]{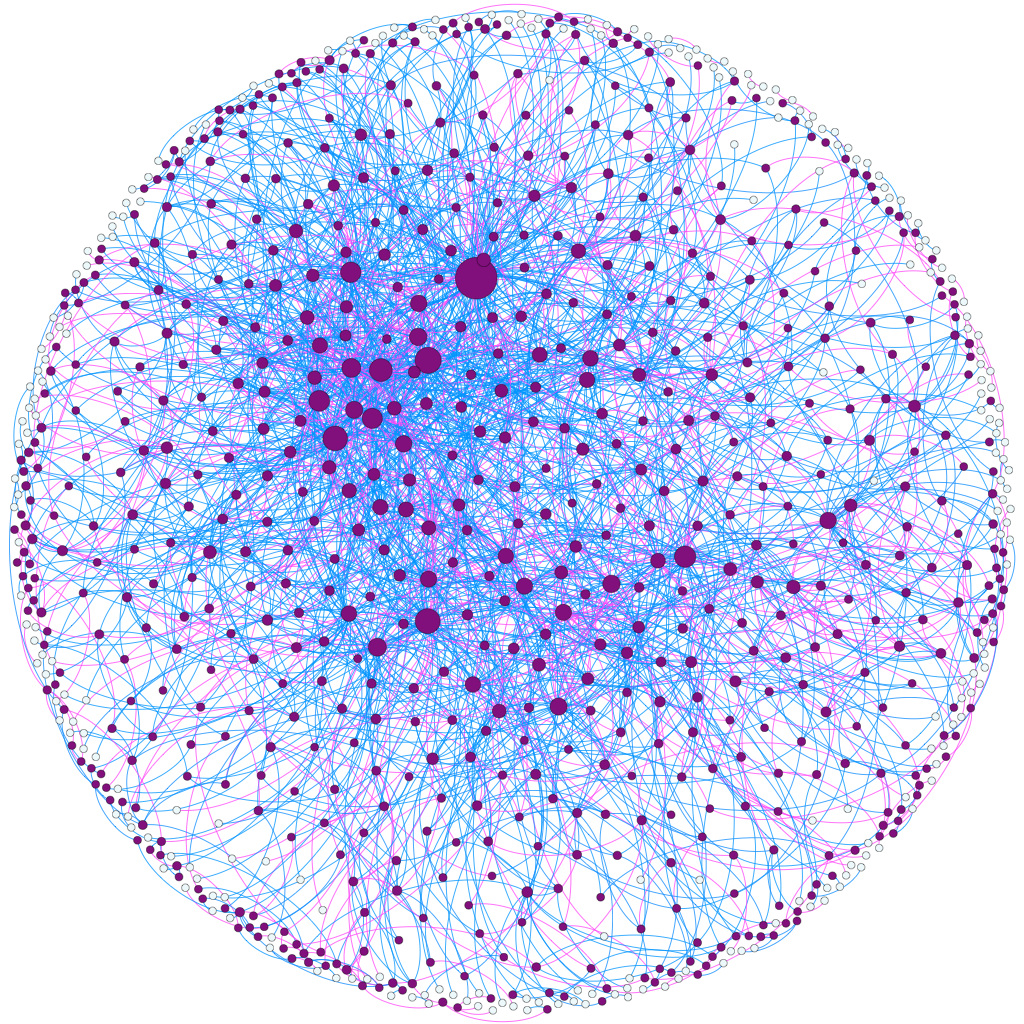}
        \caption{Social network graph for San Francisco. Blue edges are single-layer edges, while pink edges are multiplex edges. The node size is proportional to the global degree of that node.}
          \label{fig:sf}
\end{figure}

Our dataset was collected from Twitter and Foursquare in the United States between May and September 2012, where tweets and check-ins were downloaded for users who had checked-in during that time, and where those check-ins were shared on Twitter. This allows us to study the intersection of the two networks through a subset of users who have accounts and are active on both Twitter and Foursquare, and have chosen to share their check-ins to Twitter. 

\begin{table} [h!]
\centering
\small
\begin{tabular}{l*{6}{l}r}
\hline
Property & New York & Chicago &  SF &  All \\
\hline
$|V^{\cal{M}}|$ & 6,401 & 2,883 & 1,705 &  10,989\\
$|E^{T \cap F}|$ & 9,101 &  5,486 & 1,517 & 16,104 \\
$|E^{T \backslash F}|$ & 13,623 & 7,949 & 1,776 & 23,348\\
$|E^{F \backslash T}|$ & 6,394 & 4,202 & 863 & 11,459\\
$<k_{GN}>$ & 4.55 & 6.12 & 2.44  & 4.63 \\
$<k_{CN}>$ & 1.42 & 1.9 & 0.89 & 1.47\\
\hline
$tweets$ & 2,509,802 & 1,288,865 & 632,780 & 4,431,447\\
$checkins$ & 228,422 & 105,250 & 46,823 & 380,495 \\
$venues$ & 24,110 & 11,773 & 6,934 & 42,817\\
\hline
\end{tabular}
\caption{Dataset properties: number of users (nodes); number of multiplex links (edges); number of Twitter and Foursquare only edges; average global and core degrees; activity and venues per city.}
\label{tab:datt}
\end{table}

\begin{figure*}[t!]
\centering
        \begin{subfigure}[b]{0.245\textwidth}
                \includegraphics[width=\textwidth]{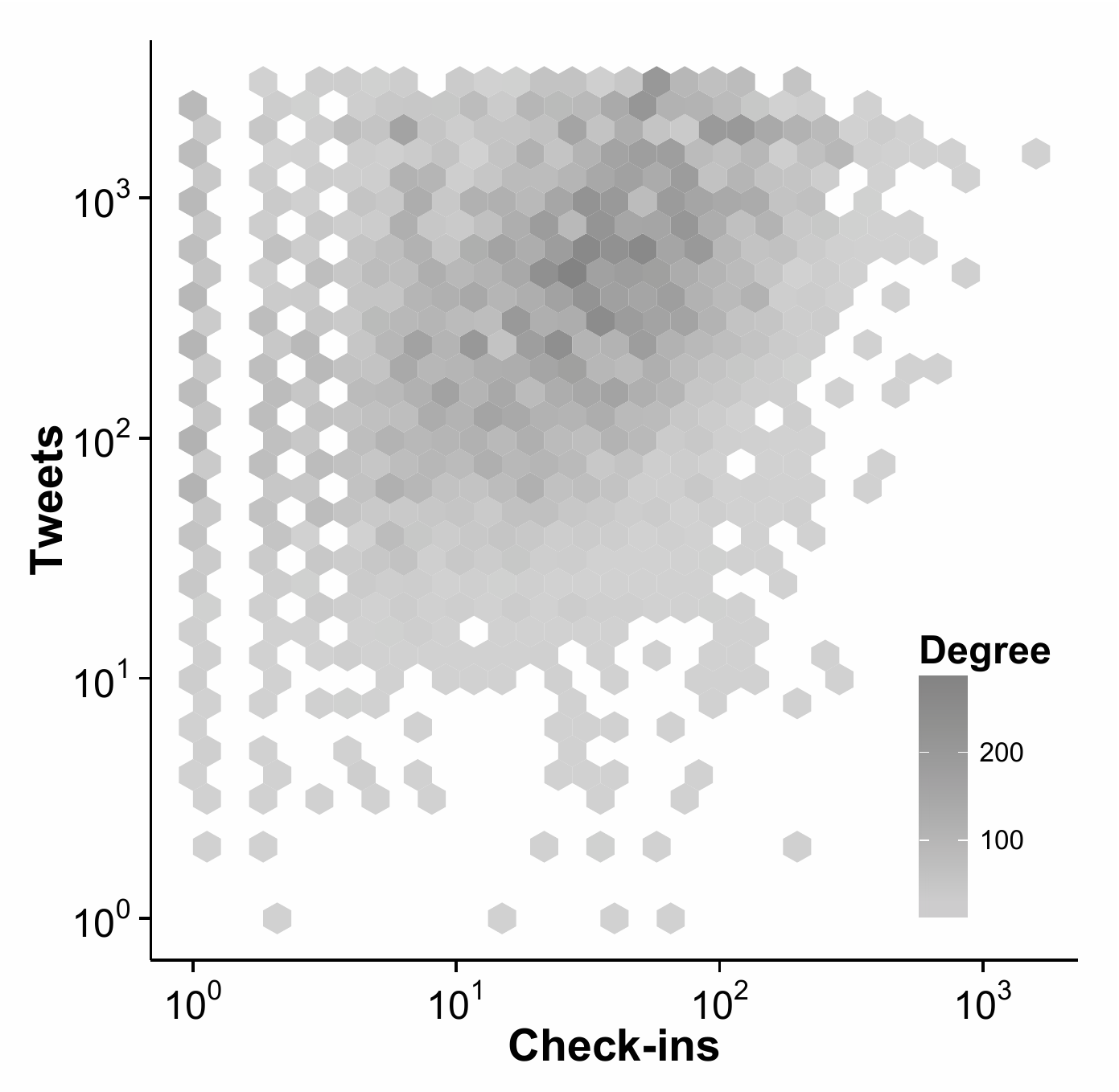}
                \caption{$k_{CNi}$ degree}
                \label{fig:ideg}
        \end{subfigure}
        \begin{subfigure}[b]{0.245\textwidth}
                \includegraphics[width=\textwidth]{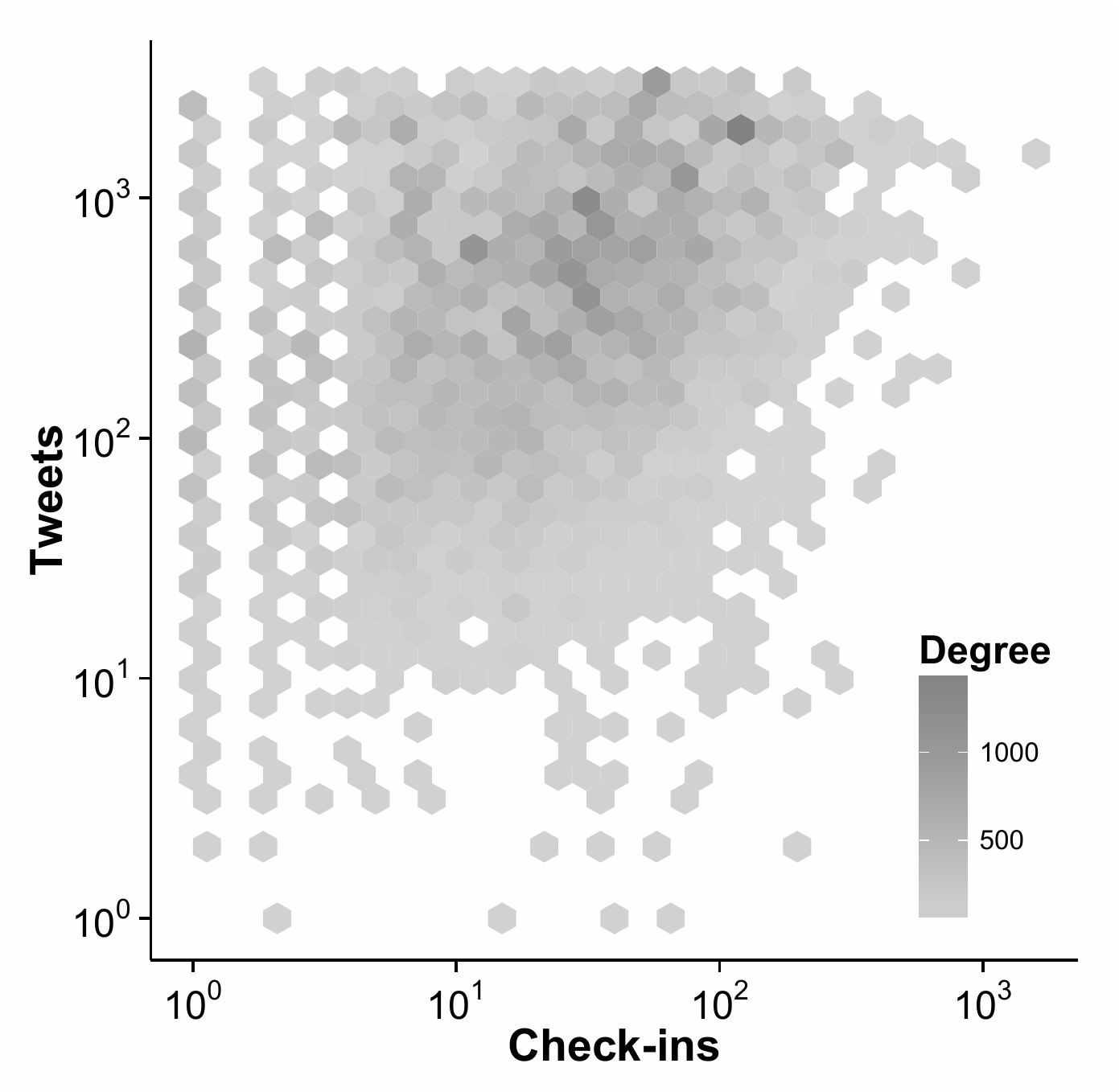}
                \caption{$k_{GNi}$ degree}
                \label{fig:udeg}
        \end{subfigure}
             \begin{subfigure}[b]{0.245\textwidth}
             \centering
                \includegraphics[width=\textwidth]{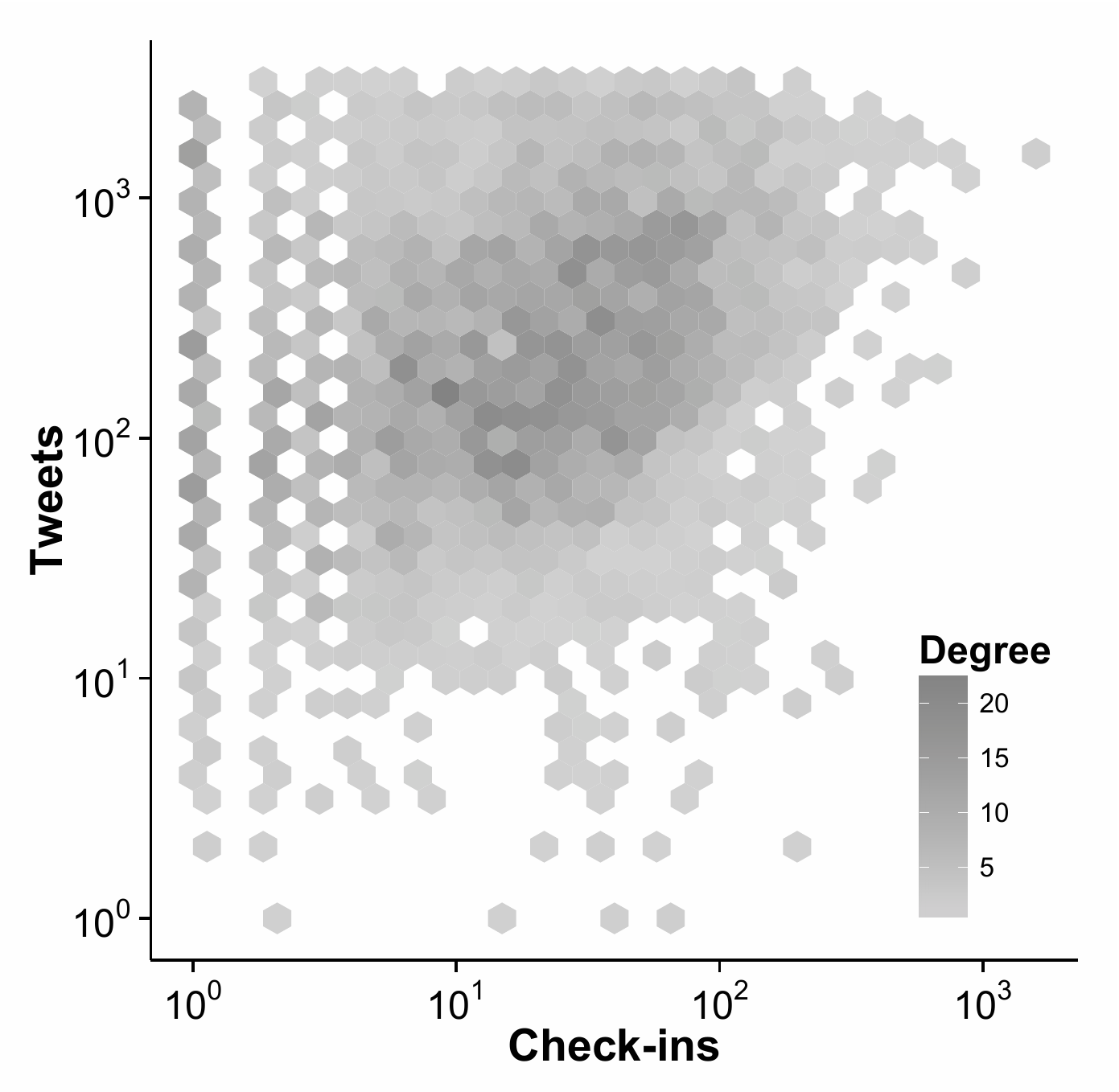}
                \caption{$mor_{ij}$ ratio}
                \label{fig:odeg}
        \end{subfigure}
                \begin{subfigure}[b]{0.245\textwidth}
                \includegraphics[width=\textwidth]{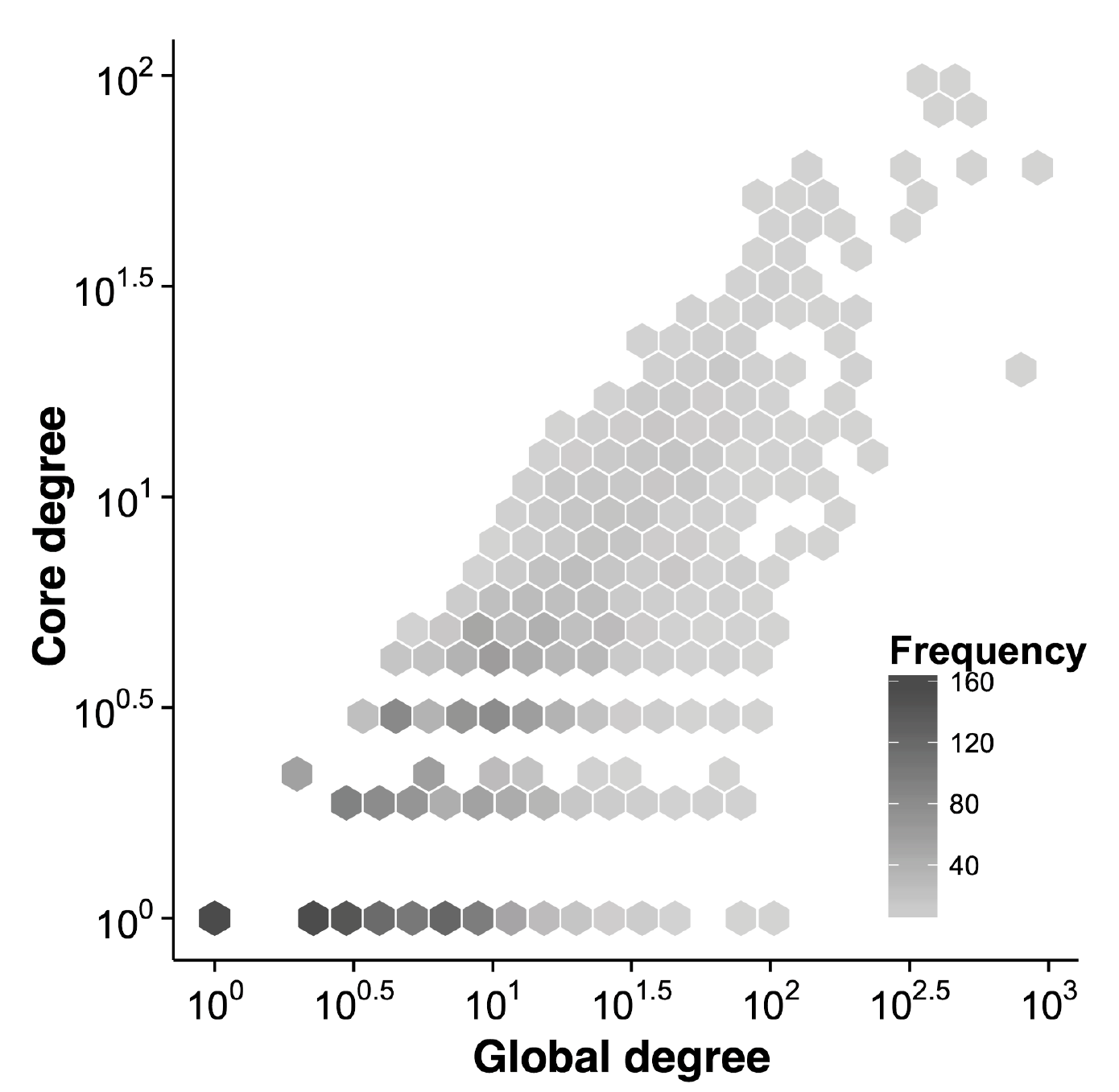}
                \caption{$k_{CNi} vs. k_{GNi}$ degree}
                \label{fig:uvi}
        \end{subfigure}
         	\centering
        \caption{Multilayer degrees of users in comparison to each other and to activity volume on both networks.}
        \label{fig:multidegs}
\end{figure*}


We focus our analysis on the top three cities in terms of activity during the period. Table~\ref{tab:datt} shows the details for each city, in terms of activity and venues, multilayer edges and degrees for each network,  where $E^{T \cap F}$ denotes the set of edges, which exist on both Twitter and Foursquare, $E^{T \backslash F}$ and $E^{F \backslash T}$ are the sets of edges on Twitter only and Foursquare only respectively.

Figure~\ref{fig:sf} additionally illustrates the case of San Francisco, where blue edges represent single-layer links on either Foursquare or Twitter, and pink edges represent multiplex links on both. We use a Fruchterman Reingold graph layout~\cite{Fruchterman1991} to show the core-periphery structure of the network, with larger nodes having a higher global degree $k_{GN}$.
In the following section, we discuss the implications of these sets in detail, where we consider all three cities together, and later evaluate each one separately.

\section{Multilayer Analysis}

We begin our analysis by exploring the intersection between the Twitter and Foursquare social networks. We observe user the degree properties across the two networks at a larger scale for all three cities, while later we perform our evaluation on each city separately.

\subsection{RQ1: Multilayer Degrees}

We introduced two degree metrics based on the multilayer neighbourhood of a node in Equations 2 and 4, where the \emph{global neighbourhood} is equivalent to the union of neighbours on both networks, and the \emph{core neighbourhood} is equivalent to the intersection of neighbours across both networks. In this section we consider how the degrees relate to user activity and each other.

In both cases (Figures~\ref{fig:ideg} and \ref{fig:udeg}), users with high activity on both networks, and in particular with high Twitter activity, have the highest degrees in both the core and global neighbourhoods. When we compare the two in Figure~\ref{fig:uvi}, we observe that their joint distribution follows the long-tail exhibited in single-layer social networks as well. Further, we observe the multiplex overlap ratio of the core to global neighbourhood degrees in Figure~\ref{fig:odeg}. This is simply the core over the global degree:

\begin{equation}
mor_{i} = \frac{k_{CNi}}{k_{GNi}}
\end{equation} 

which indicates the percent of multiplex links in $i$'s multilayer neighbourhood. High activity nodes across both layers at the centre of Figure~\ref{fig:odeg} have the highest overlap. 

In Figure~\ref{fig:uvi}, we compare the two multilayer degrees. We note that the majority of users have a low degree in both, and there is a relationship between the two. The core degree is bound by the global degree and is always a fraction of it, while the global degree may never exceed the sum of the individual layer degrees. This relationship is apparent in the figure, where \emph{the highest degree users are those who have a large number of links which overlap (multiplex links)}. This can be due to the fact that these users are more engaged across the two platforms. We further explore the value of link multiplexity in the following section.

\subsection{RQ2: Link Multiplexity}


\begin{figure*}[t!]
               \centering
        \begin{subfigure}[b]{0.24\textwidth}
                \includegraphics[width=\textwidth]{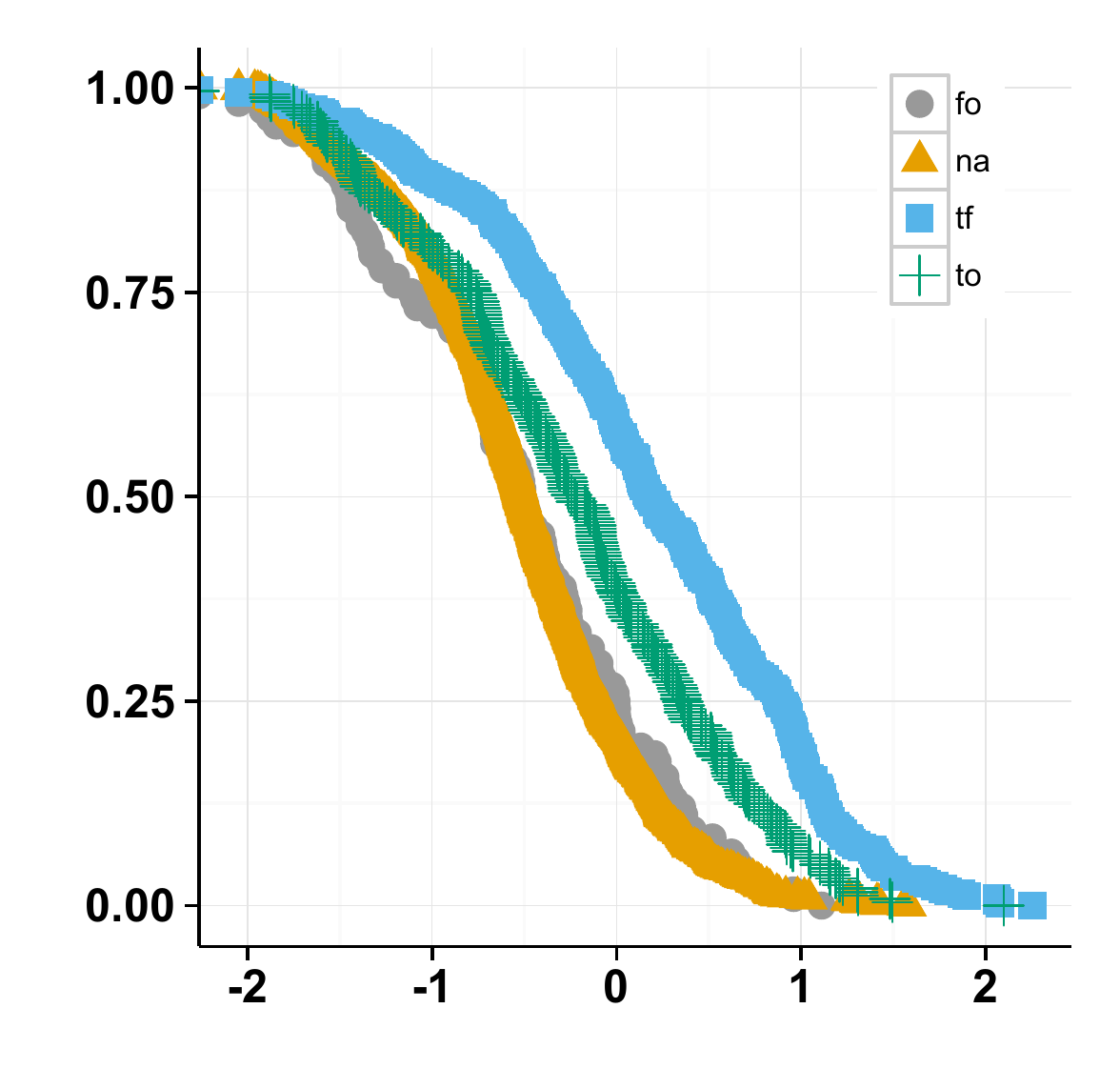}
                \caption{Twitter Adamic/Adar overlap}
                \label{fig:aat}
        \end{subfigure}%
        \begin{subfigure}[b]{0.24\textwidth}
                \includegraphics[width=\textwidth]{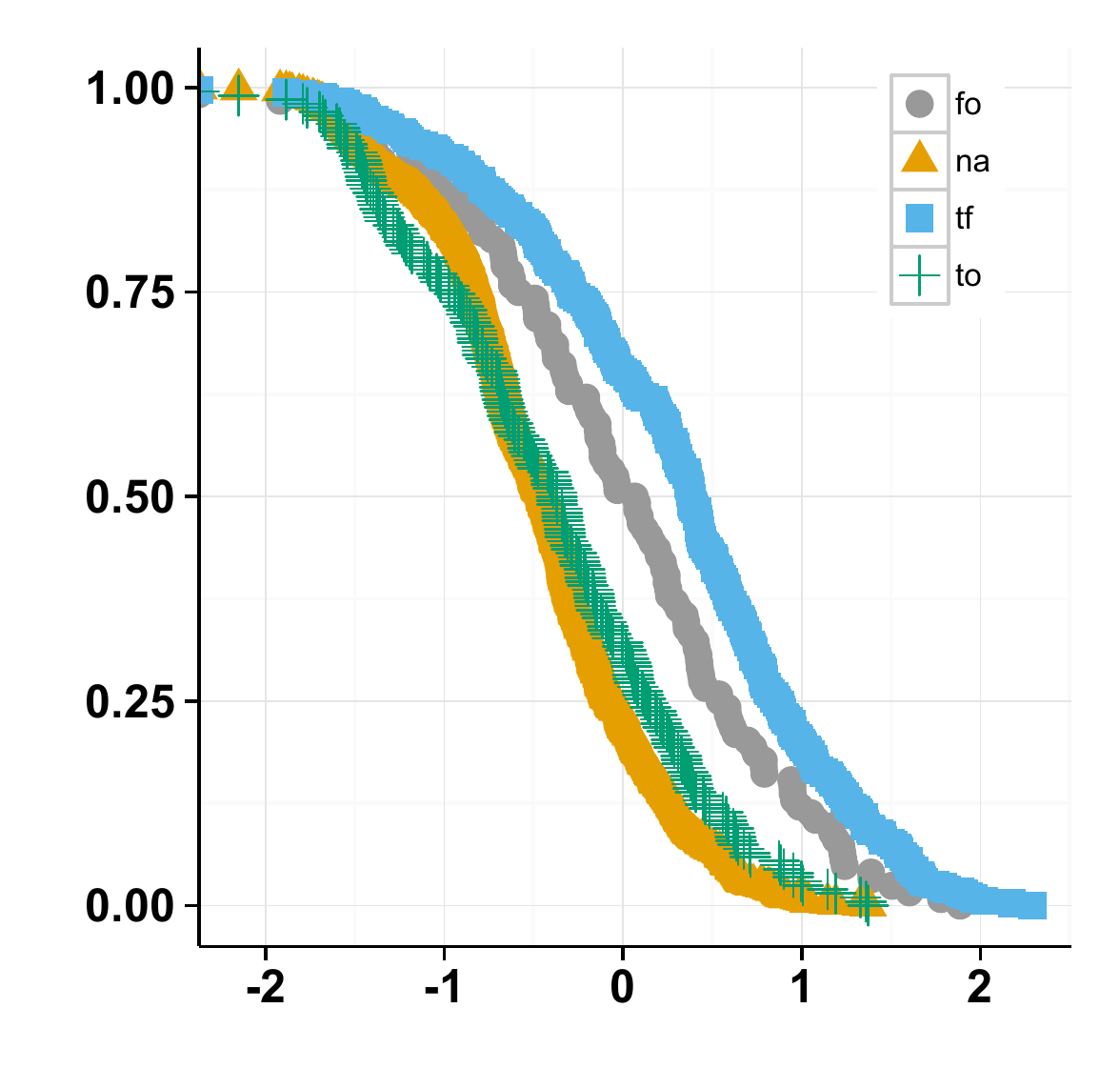}
                \caption{Foursq. Adamic/Adar overlap}
                \label{fig:aaf}
        \end{subfigure}
                \begin{subfigure}[b]{0.24\textwidth}
                \includegraphics[width=\textwidth]{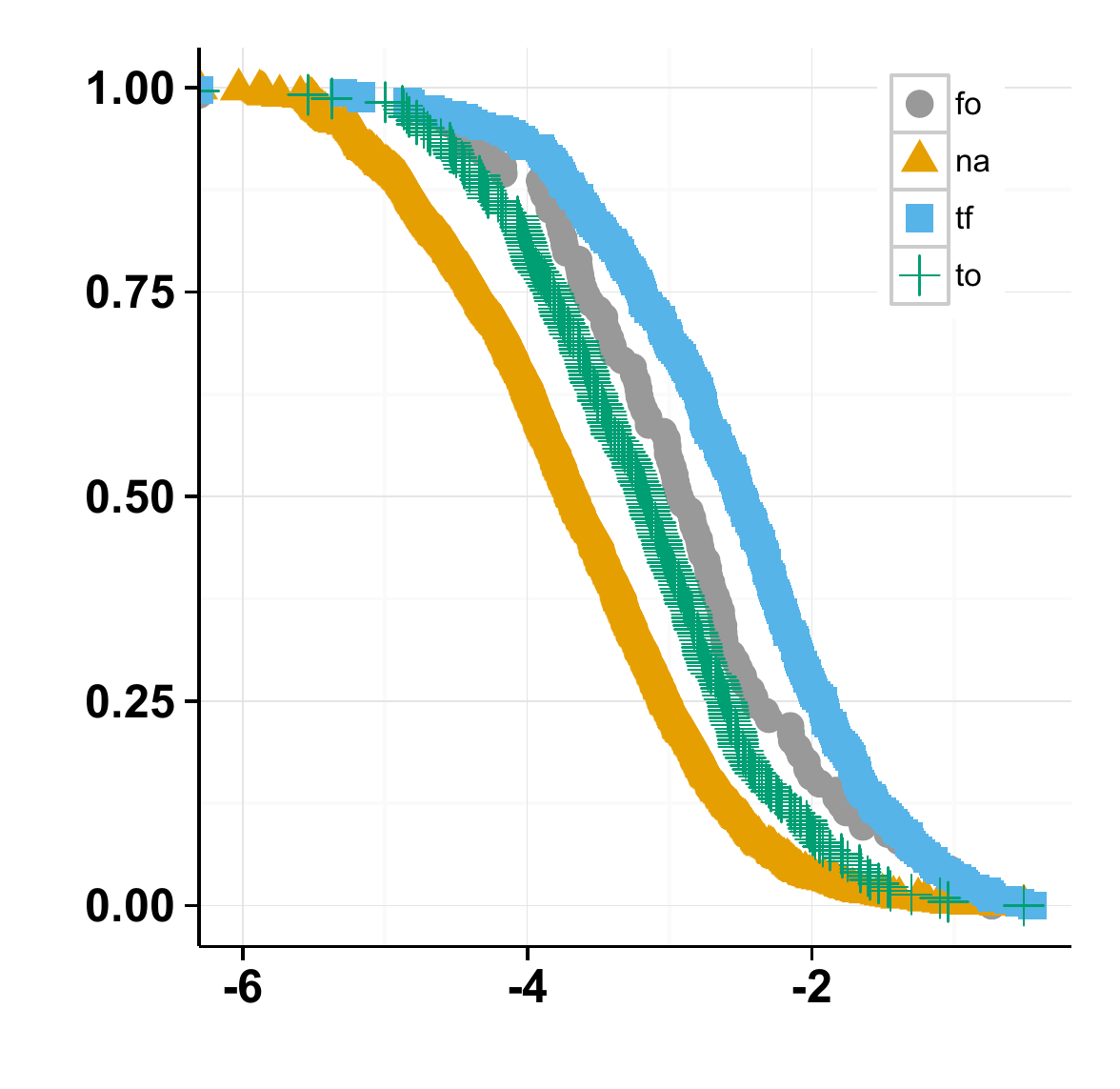}
                \caption{$\Gamma_{GNij}$ Adamic/Adar overlap}
                \label{fig:aau}
        \end{subfigure}
                \begin{subfigure}[b]{0.24\textwidth}
                \includegraphics[width=\textwidth]{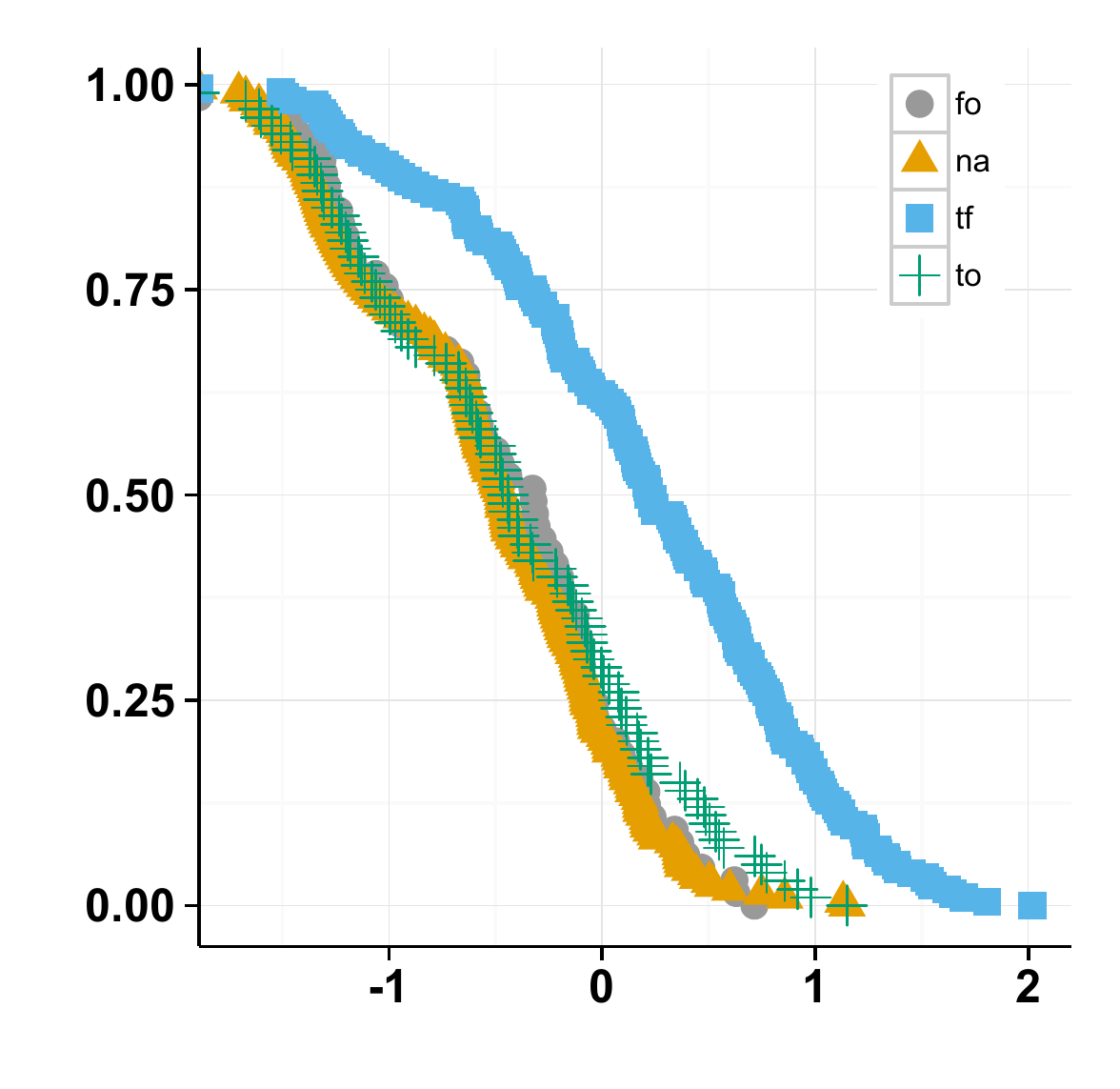}
                \caption{$\Gamma_{CNij}$ Adamic/Adar overlap}
                \label{fig:aai}
        \end{subfigure}
        \caption{CCDF function of the log Adamic/Adar metric for the different neighbourhoods between the four link types.}
        \label{fig:structure}
\end{figure*}

We study the three types of links as described in our multilayer model above: multiplex links on both Twitter and Foursquare, which we denote as  \emph{tf} for simplicity; single-layer links on Foursquare only (denoted as \emph{fo}); single-layer links on Twitter only (denoted as \emph{to}), and compare these to unconnected pairs of users (denoted as \emph{na}). We consider reciprocal Twitter links only, where $e_{ij}, e_{ji} \in E^T$. Reciprocal relationships in Twitter have been considered as equivalent to undirected ones in other OSNs~\cite{Kwak2010}. 

\subsection{Multiplexity and Neighbourhoods}

The number of common friends has been shown to be an important indicator of a link in social networks~\cite{liben2007link}. Moreover, the neighbourhood overlap weighted on the popularity of common links between two users has been shown to be a good predictor of friendship in online networks~\cite{adamic2001}. Figure~\ref{fig:structure} shows the Adamic/Adar metric of neighbourhood similarity across the various single and multilayer neighbourhoods described in Section 3, and the four link types. 

The Adamic/Adar metric is distinctly higher for multiplex links. In agreement with previous studies of tie strength~\cite{Gilbert2009}, we observe that multiplex links share a greater overlap in all single and multilayer neighbourhoods. In single-layer neighbourhoods (Figure~\ref{fig:aat} and \ref{fig:aaf}) we observe that after multiplex links, those links internal to the network under consideration have a higher overlap than exogenous ones (\emph{to} in Figure~\ref{fig:aat} and \emph{fo} in Figure~\ref{fig:aaf}), followed by unconnected pairs, which have the least overlap.

With respect to the multilayer neighbourhoods, we can observe a much more pronounced overlap across the link types. While the global neighbourhood overlap follows a similar distribution to the single-layer neighbourhoods but at a much lower scale, in Figure~\ref{fig:aai} we can observe more clearly that unconnected pairs share little if any neighbours, while multiplex links have a significant overlap. With respect to the global neighbourhood (Figure~\ref{fig:aau}), both Foursquare only and Twitter only links share significantly more overlap (scale is higher on x axis) than when observing the single-layer neighbourhoods in Figures~\ref{fig:aat} and \ref{fig:aaf}. This indicates that some common neighbours lie across layers, and not just within, \emph{the global neighbourhood revealing a more complete image of connectivity, which stretches beyond the single network}. 

The core neighbourhood overlap is most prominent for multiplex links (Figure~\ref{fig:aai}), which indicates that they share more friends across networks than any other type of link. While this is expected, it \emph{confirms that the neighbourhood overlap is a good indicator of multiplexity in ties}, and is particularly strengthened in its weighted form through the Adamic/Adar metric of neighbourhood similarity.

\subsection{Multiplexity and Interaction}

The volume of interactions between users is often used as a measure of tie strength~\cite{onnela2007}. In this section we compare how the volume of interactions reflects on multiplex and single-layer links. We consider the following interactions on Twitter and Foursquare:\\
\textbf{Number of mentions:} This interaction feature simply measures the number of times user $i$ has mentioned user $j$ on Twitter during the period. Any user on Twitter can mention any other user and need not have a directed or undirected link to the user he is mentioning.\\
\textbf{Number of common hashtags:} Similarity between users on Twitter can be captured through common interests. Topics are commonly expressed on Twitter with hashtags using the \# symbol. Similar individuals have been shown to have a greater likelihood of forming a tie through the principles of homophily~\cite{mcpherson2001}.\\
\textbf{Number of colocations:} The number of times two users have checked-in to the same venue within a given time window. In order to reduce false positives, we consider a shorter time window of 1 hour only. Two users at the same place, at the same time on multiple occasions, increases the likelihood of them knowing each other (and having a link on social media). We weight each colocation on the popularity of a place in terms of the total user visits, to reduce the probability that colocation is by chance at a large hub venue such an airport or train station. \\
\textbf{Distance:}  Human mobility and distance play an important role in the formation of links, both online and offline, and have been shown to be highly indicative of social ties and useful for link prediction~\cite{Wang2011}. We calculate the distance between the geographic coordinates of two users' most frequent check-in locations as the Haversine distance, the most common measure of great-circle spherical distance:
\begin{equation}
dist_{ij} = haversine(lat_i,lon_i, lat_j,lon_j)
\end{equation} 

where the coordinate pairs for $i,j$ are of the places where those users have checked-in most frequently, equivalent to the mode in the multiset of venues where they have checked-in. We only consider users who have more than two check-ins over the whole period, and resolve ties by picking an arbitrary venue location from the top ranked venues of a user.

\begin{figure}[t!]
       \centering
        \begin{subfigure}[b]{0.22\textwidth}
                \includegraphics[width=\textwidth]{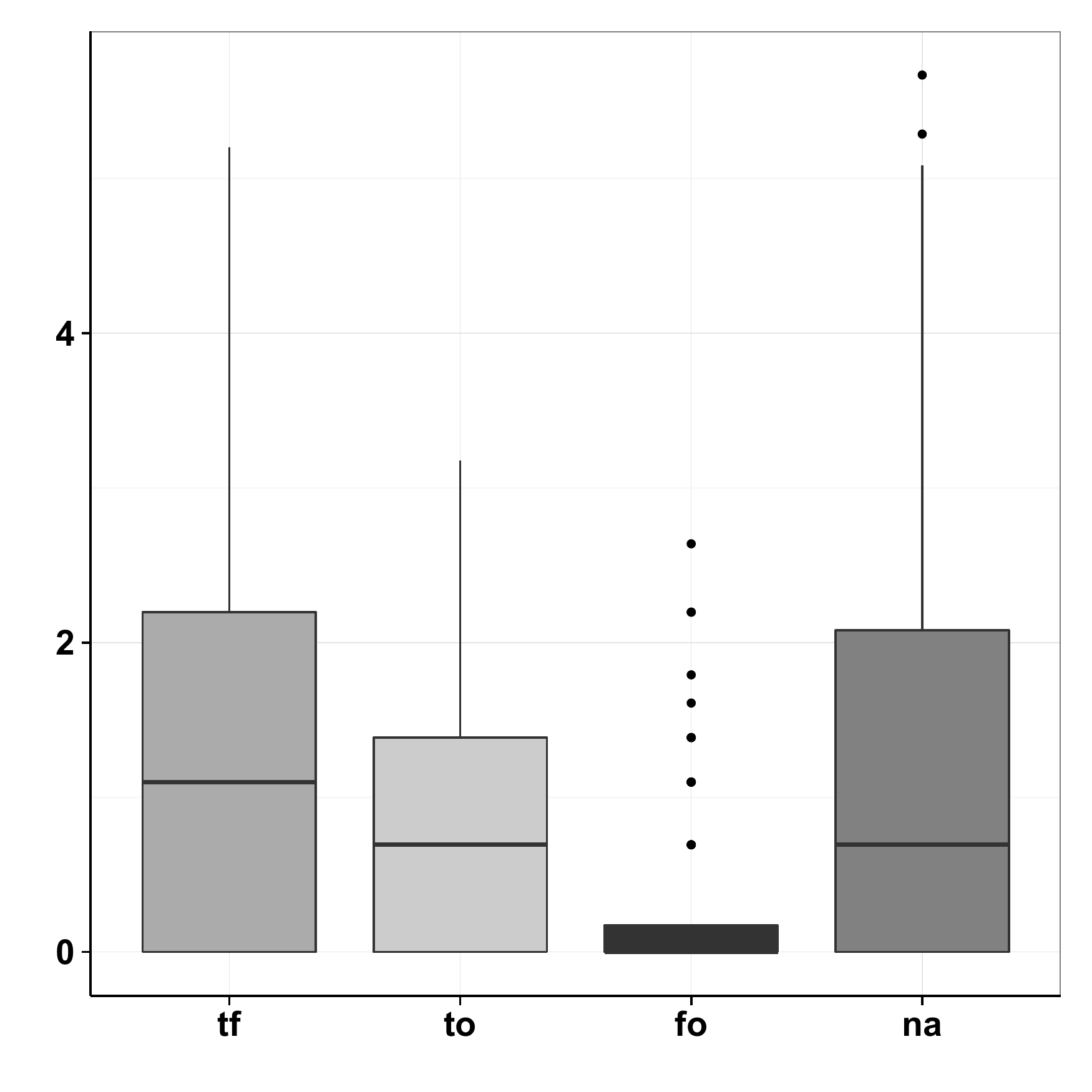}
                \caption{Log scale mentions}
                \label{fig:men}
        \end{subfigure}%
               \centering
        \begin{subfigure}[b]{0.22\textwidth}
                \includegraphics[width=\textwidth]{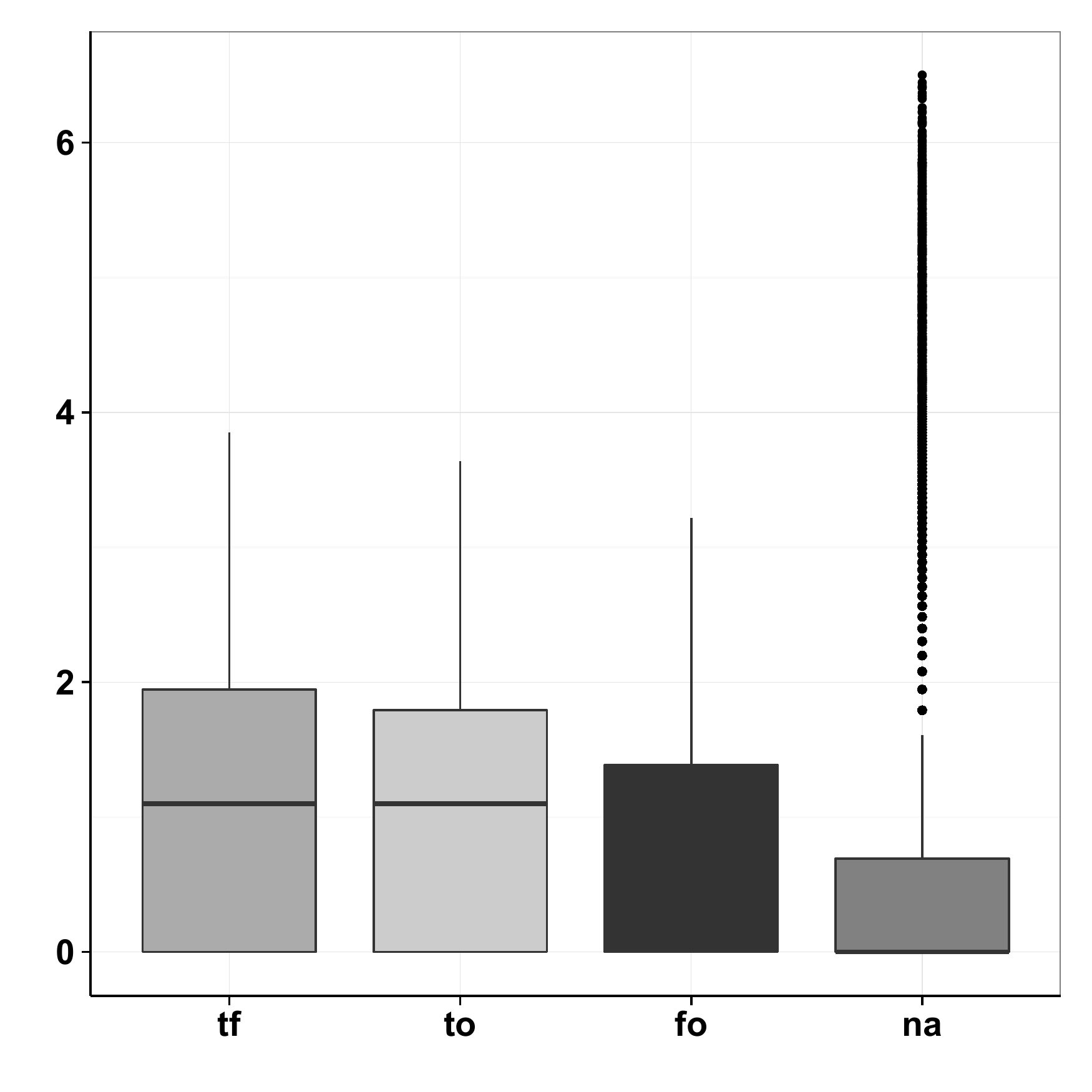}
                \caption{Log scale hashtags}
                \label{fig:hash}
        \end{subfigure}%
        
        \centering
        \begin{subfigure}[b]{0.22\textwidth}
                \includegraphics[width=\textwidth]{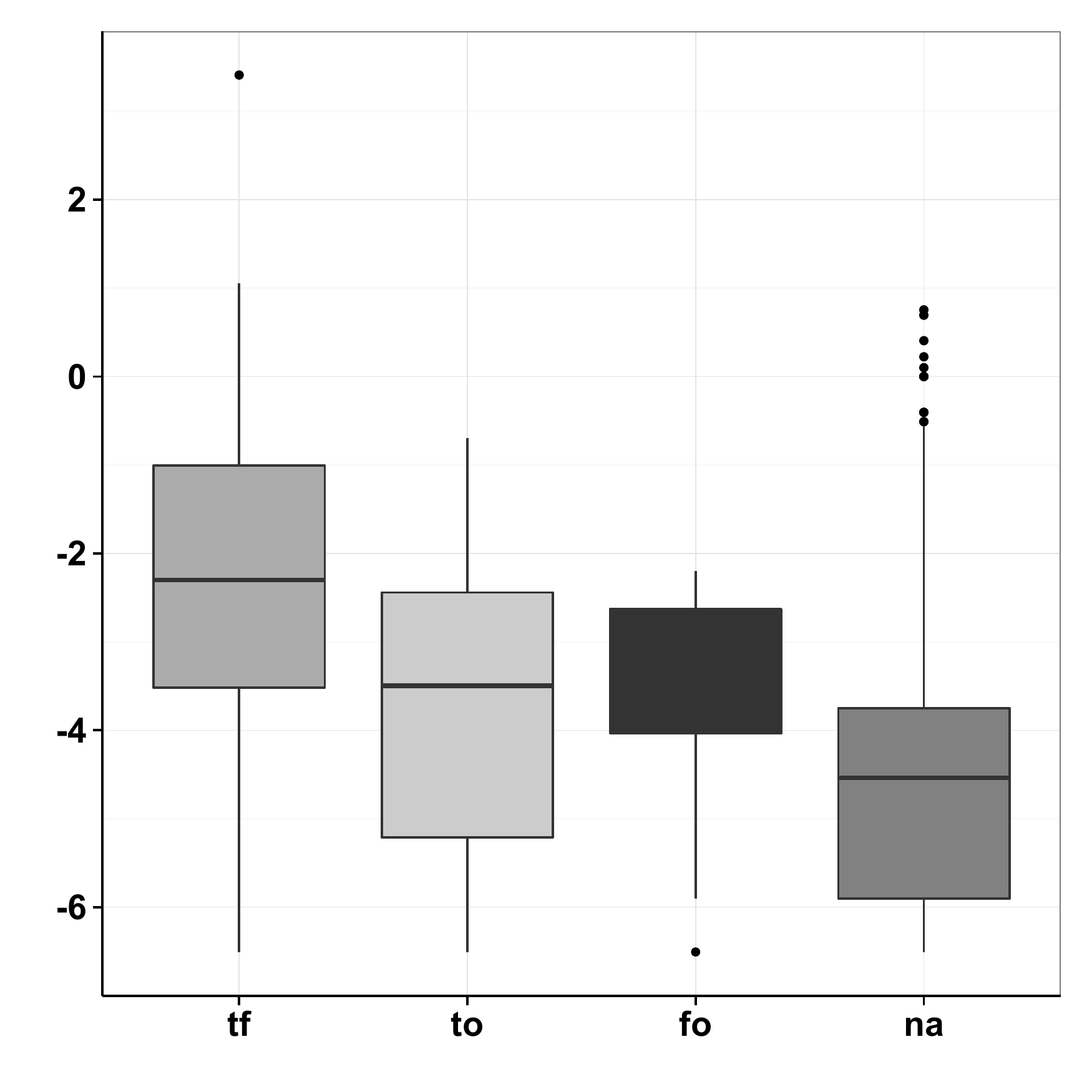}
                \caption{Log scale colocations}
                \label{fig:coloc}
        \end{subfigure}%
        \begin{subfigure}[b]{0.22\textwidth}
                \includegraphics[width=\textwidth]{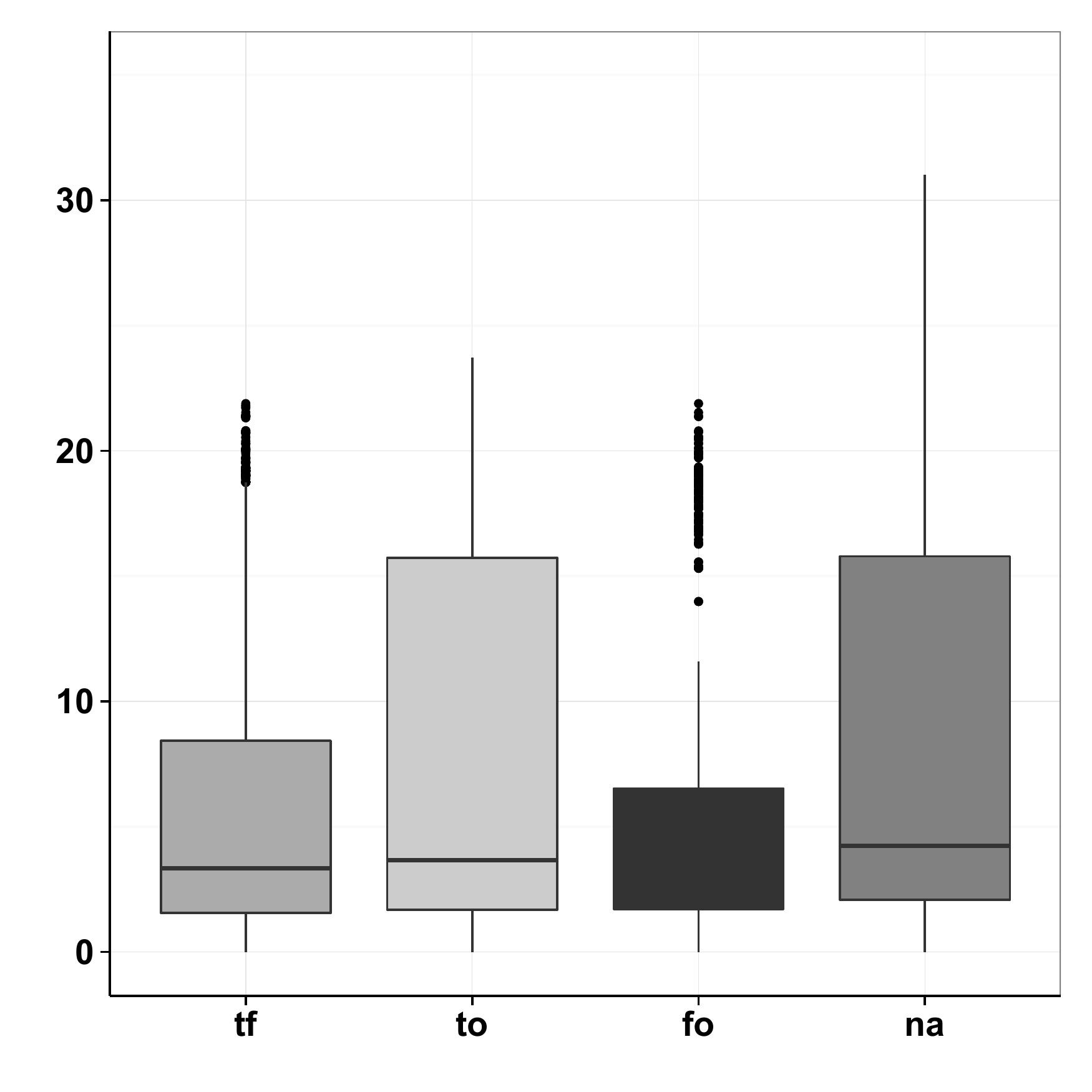}
                \caption{Distance in km}
                \label{fig:dist}
        \end{subfigure}
        
             \centering
        \begin{subfigure}[b]{0.22\textwidth}
                \includegraphics[width=\textwidth]{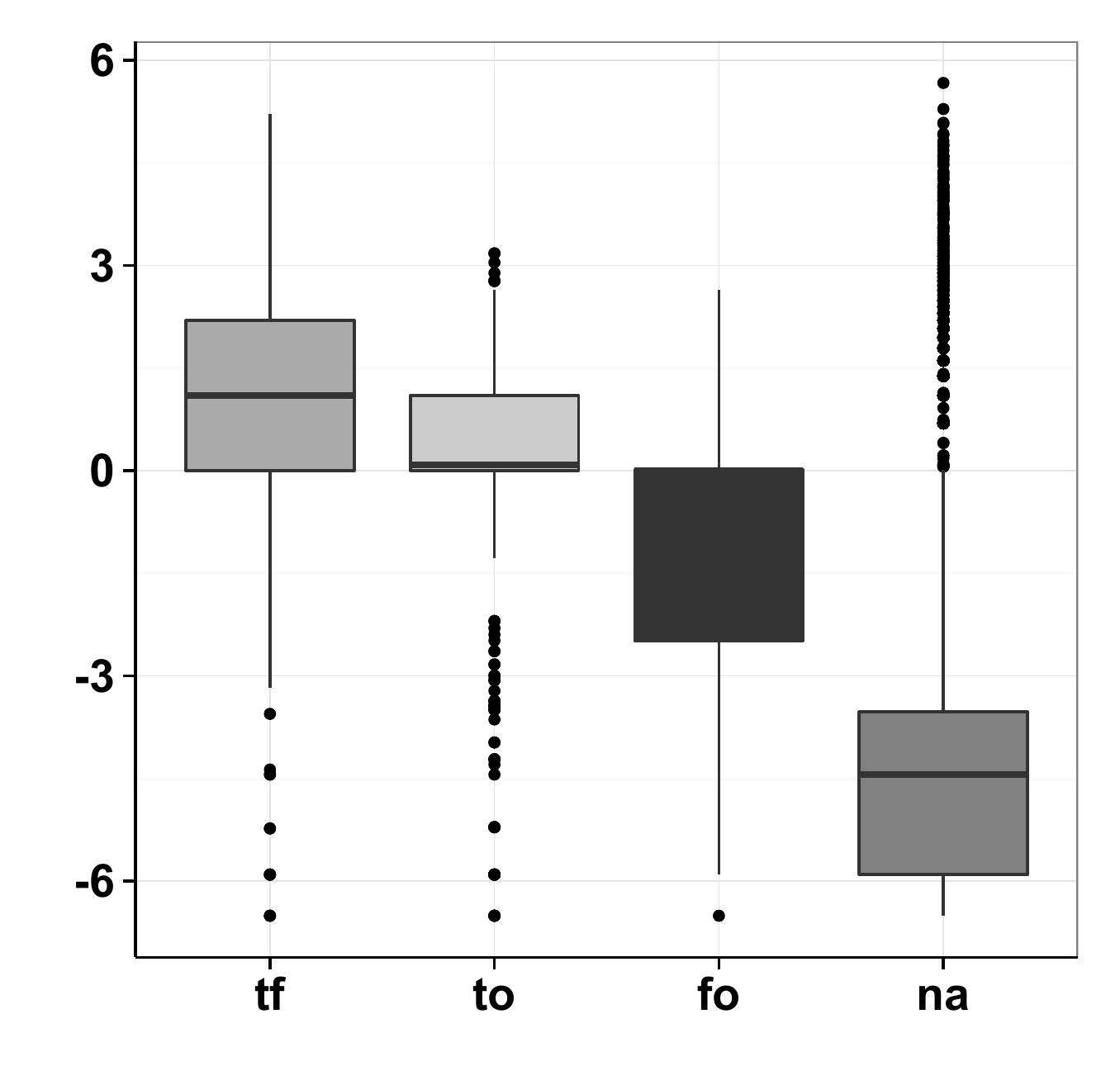}
                \caption{Colocations and mentions}
                \label{fig:mult1}
        \end{subfigure}%
        \begin{subfigure}[b]{0.22\textwidth}
                \includegraphics[width=\textwidth]{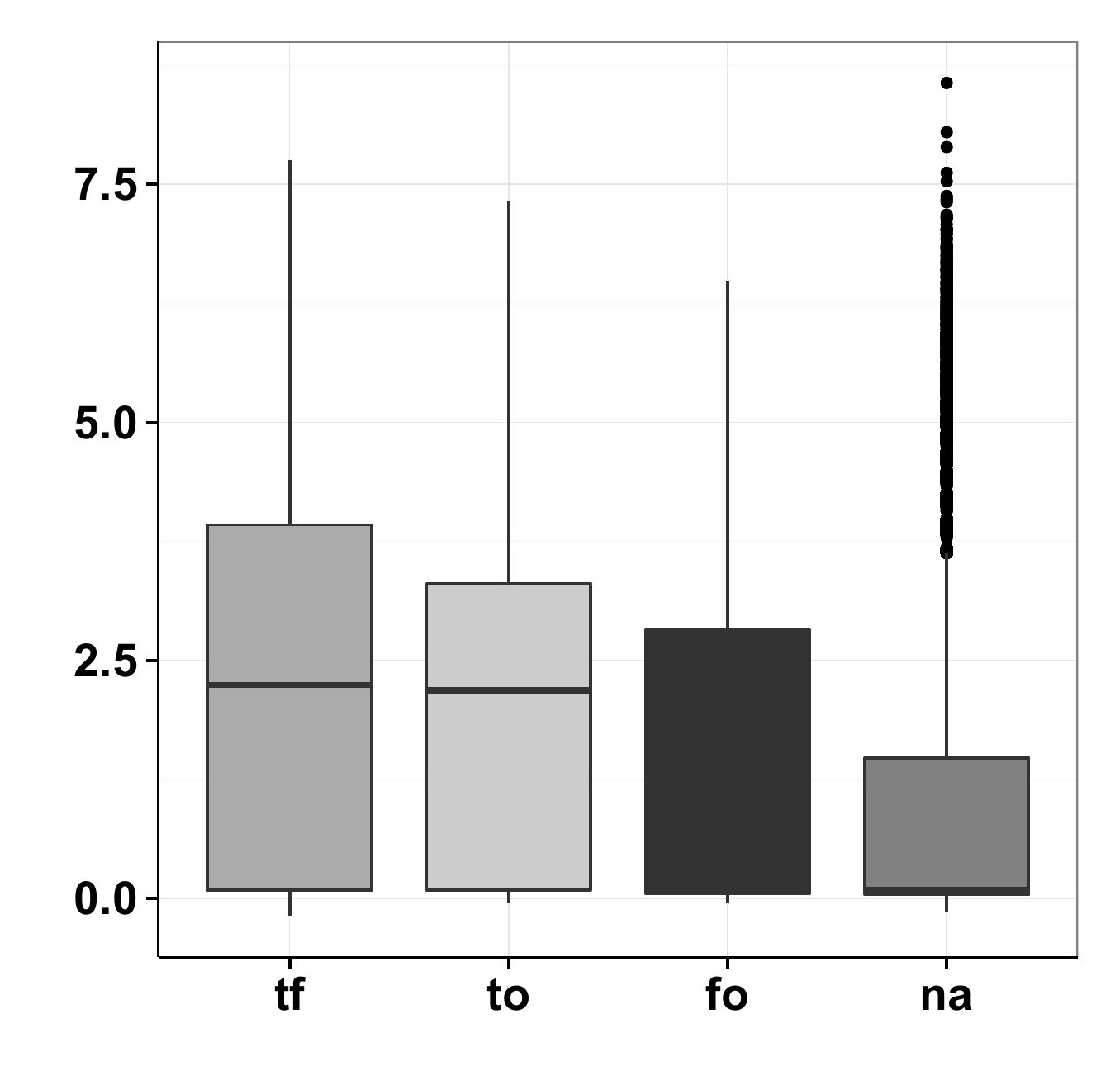}
                \caption{Hashtags over distance}
                \label{fig:mult2}
        \end{subfigure}
        \caption{Interaction features for the different link types.}
        \label{fig:interactions}
\end{figure}

In Figures~\ref{fig:men} to \ref{fig:dist}, we observe four types of geographic and social interaction on the two social networking services, where each box-and-whiskers plot represents an interaction between multiplex links (\emph{tf}), Twitter only (\emph{to}), Foursquare only (\emph{fo}), and unconnected pairs (\emph{na}) on the x axis. On the y axis we can observe the distribution in four quartiles, representing 25\% of values each. The dark line in the middle of the box represents the median of the distribution, while the dots are the outliers. The ``whiskers" represent the top and bottom quartiles, while the boxes are the middle quartiles of the distribution.

In terms of Twitter mentions (Figure~\ref{fig:men}), multiplex ties and non-connected pairs of users exhibit  an overall greater number of mentions than any other group, including the Twitter only group. It is uncommon that pairs connected on Foursquare only mention each other. Mentions are quite common between users who are not connected on any network, which may be as a result of mentioning celebrities and other commercial accounts. This is not  the case for hashtags, where we find that almost all of unconnected users share 10 or less hashtags with the exception of outliers. 
Hashtags distinguish the link type between users better than mentions.

With regards to Foursquare interaction, multiplex ties have the highest probability of multiple colocations, with Foursquare and Twitter only ties having less, and unconnected pairs more so with the exception of some outliers. In terms of distance, Twitter only and unconnected pairs are the furthest apart in terms of most frequented location, making multiplex and Foursquare links more distinguishable through this feature, as those pairs have less distance between their most frequented locations. 

Although \emph{there is certainly greater interaction between multiplex links}, followed by Twitter only and Foursquare only links, we would like to eliminate the randomness introduced by the positive results for unconnected pairs (\emph{na}).  We propose two multilayer interaction metrics combining heterogenous features from both networks in order to  better distinguish between the different link types. Firstly, we define the global similarity as the Twitter similarity over Foursquare distance as: 

\begin{equation}
sim_{GNij} = \frac{sim_{ij}^a}{dist_{ij}^b}
\end{equation}

where $sim$ can be replaced with any type of similarity, which is the mass or sum of that similarity for a pair of users, and $a,b$ are exponents which can be tuned to optimise the features. Figure~\ref{fig:mult2} shows how this feature captures the different levels of links (a=2, b=1). We additionally frame a feature which captures the complete interaction across layers of social networks: 

\begin{equation}
int_{GNij} = \sum_\alpha^M |int^\alpha_{ij}|
\end{equation}

where $int$ can be any type of interaction of layer $\alpha$, this can be further refined by giving a weight to each interaction but in our case, we consider the coefficient to be equal to 1 and use colocations from the Foursquare layer and mentions from the Twitter layer to express the global interaction of two users in the multilayer network. This feature allows us to capture the levels of different link types significantly better as shown in Figure~\ref{fig:mult1}. 

Although we base our analysis on only  two of many possible communication channels online, we are nonetheless able to observe the greater overlap of neighbourhoods and higher intensity of interaction characteristic of multiplex links, which is in consistency with the theory of media multiplexity~\cite{hay2005}.  We evaluate the predictive performance of the union of the features presented in the following section.

\section{Multiplexity \& Link Prediction}

In this section we address the link prediction problem across layers of social networks, and aim to answer our final two research questions: \emph{Can we predict one network using information from the other?}, and \emph{Can we predict multiplex links in OSNs?} We evaluate the likelihood of forming a social tie as a process that depends on a union of factors, using the Foursquare, Twitter, and multilayer features we have defined up until now in a supervised learning approach, and comparing their predictive power in terms of AUC scores for each feature set respectively. 

\begin{figure*}[t!]
	 \begin{subfigure}[b]{0.19\textwidth}
                \includegraphics[width=\textwidth]{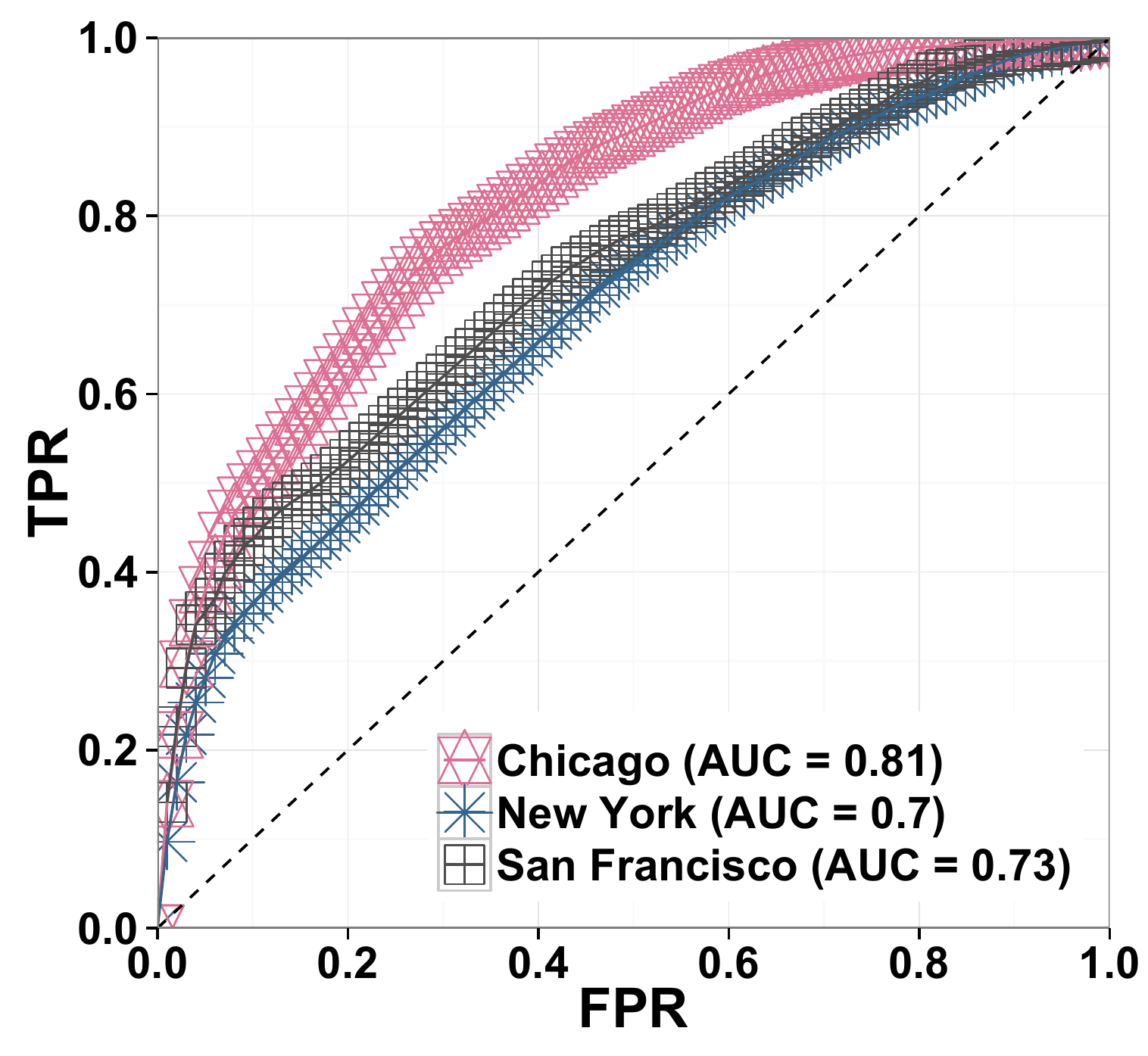}
                \caption{Foursq. link prediction}
                \label{fig:roc_fsq}
        \end{subfigure}%
         \begin{subfigure}[b]{0.19\textwidth}
                \includegraphics[width=\textwidth]{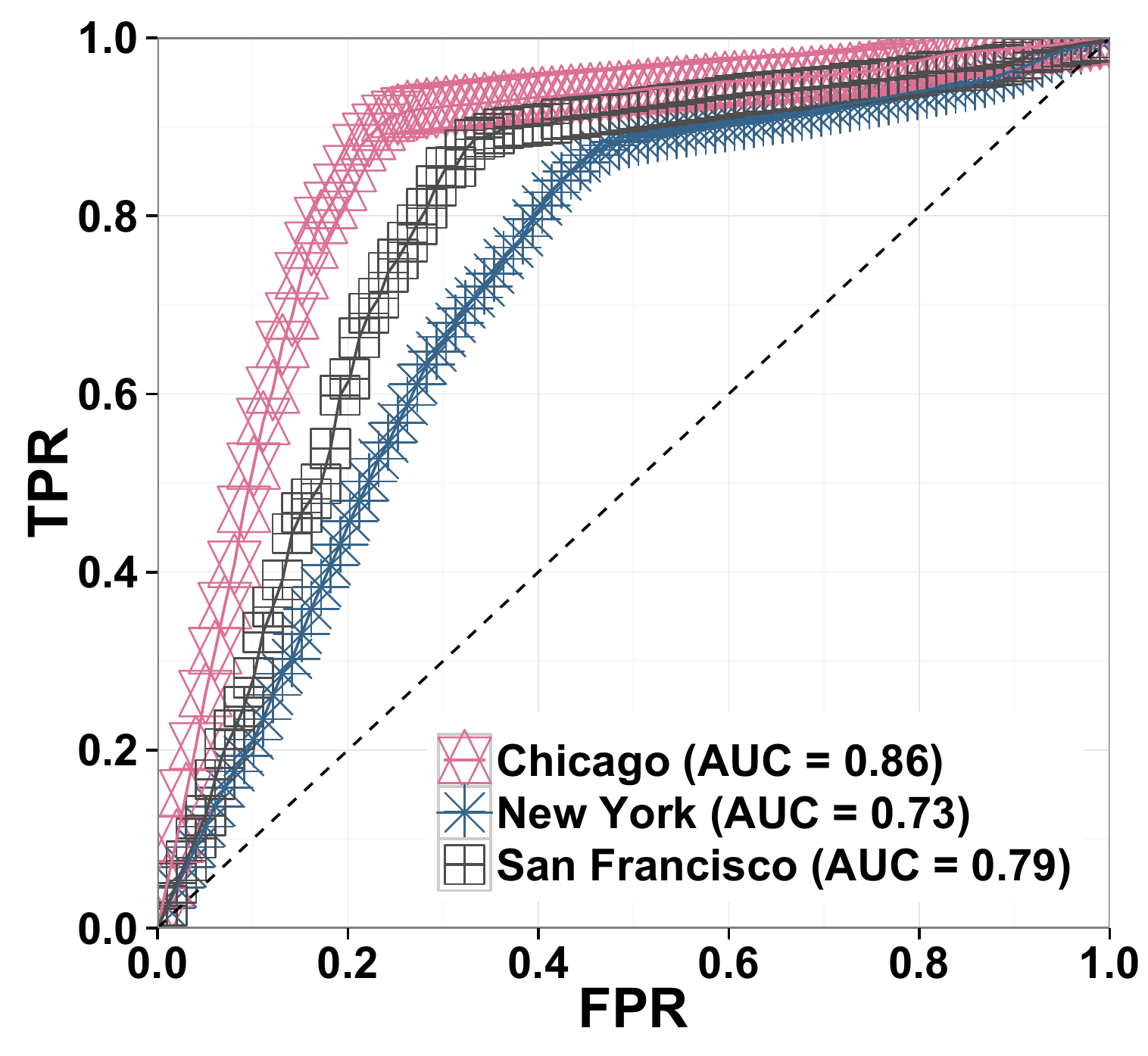}
                \caption{Twitter link prediction}
                \label{fig:roc_mfsq0}
          \end{subfigure}
	\centering
	 \begin{subfigure}[b]{0.19\textwidth}
                \includegraphics[width=\textwidth]{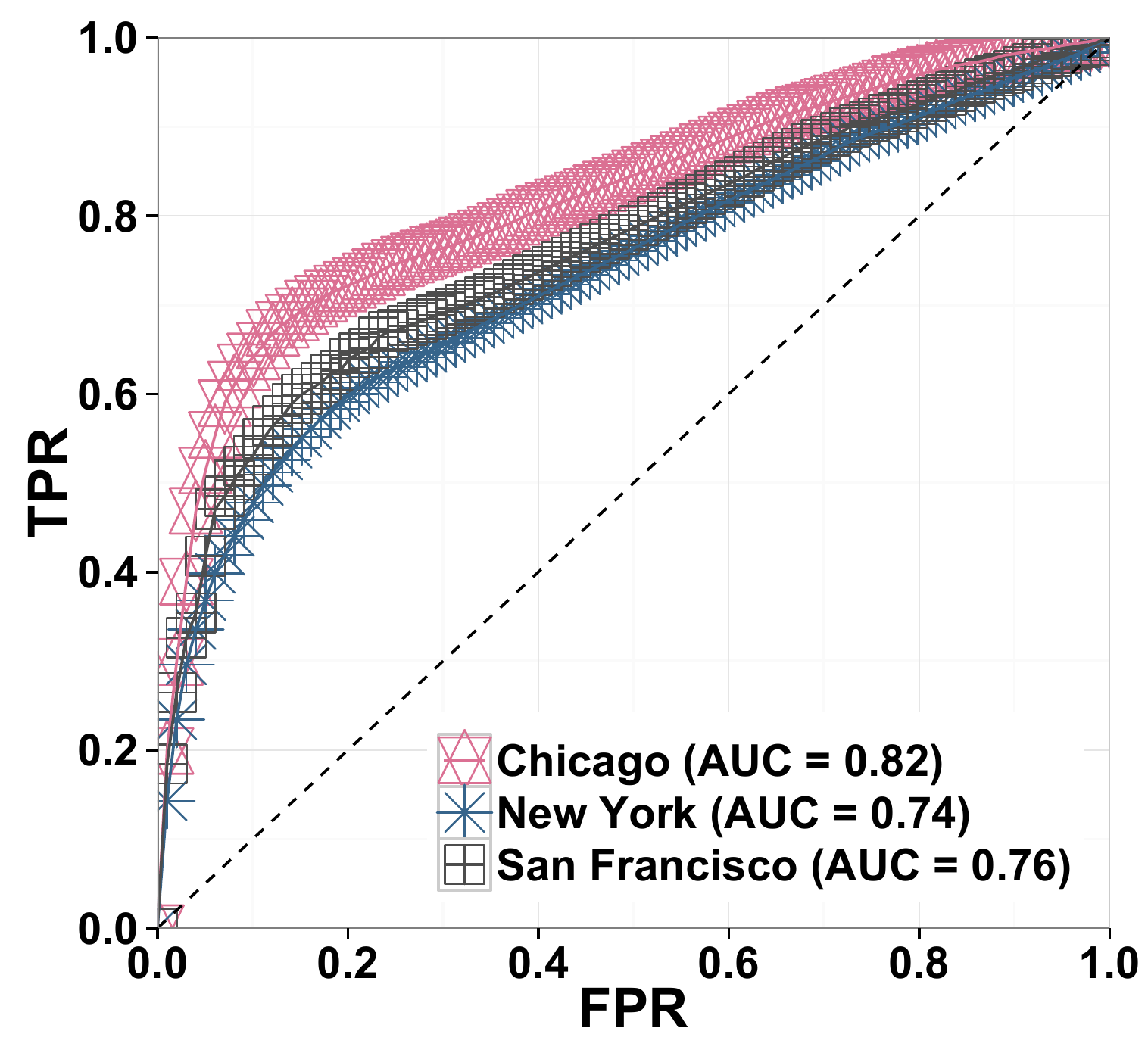}
                \caption{Twitter feature set}
                \label{fig:m_fsq}
        \end{subfigure}%
         \begin{subfigure}[b]{0.19\textwidth}
                \includegraphics[width=\textwidth]{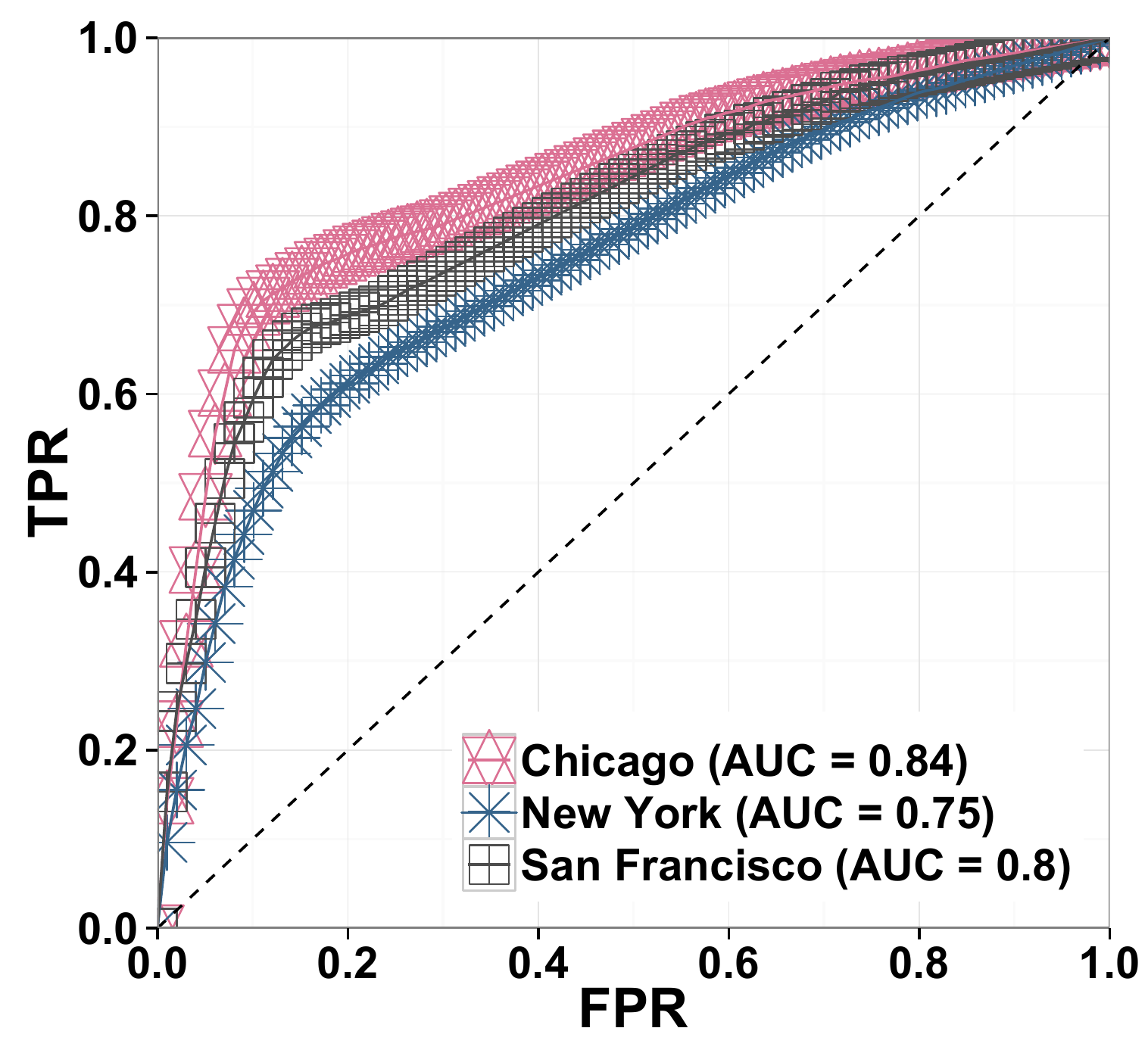}
                \caption{Foursquare feature set}
                \label{fig:rm_twt}
          \end{subfigure}
              \begin{subfigure}[b]{0.19\textwidth}
              \centering
                \includegraphics[width=\textwidth]{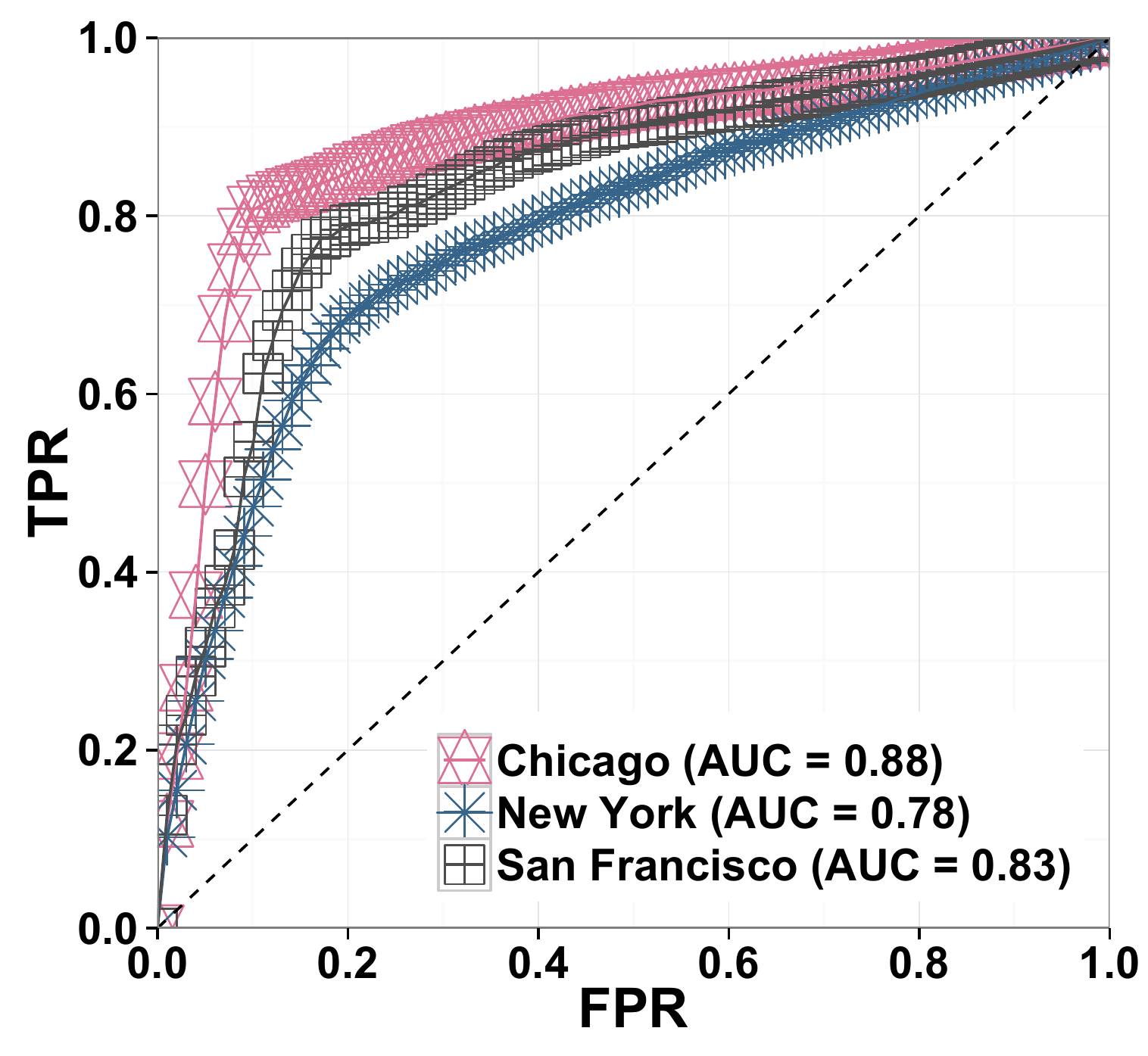}
                \caption{Multilayer feature set}
                \label{fig:roc_mfsq}
          \end{subfigure}
  \caption{ROC curves for the Random Forest classifier and Area Under the Curve (AUC) scores for each city dataset.}
    \label{fig:roc1}
\end{figure*}

\subsection{Prediction Space}

The main motivation for considering multiple social networks in a multilayer construct is that each layer carries with it additional information about the links between the same users, which can potentially enhance the predictive model. In light of the multilayer nature of OSNs, we are also interested in whether we can achieve better prediction by combining features from multiple networks.

Formally, for two users $i,j \in \cal{M}$, where $V^{\cal{M}}$ are the nodes (users) that are present in any layer of the multilayer network, we employ a set of features that output a score $r_{ij}^\alpha$ so that all possible pairs $V^{\cal{M}} \times V^{\cal{M}}$ are ranked according to their expectation of having a link $e_{ij}^\alpha$ on a specific layer $\alpha$ in the network.
We specify and evaluate two distinct prediction tasks. 

Our first goal is to rank pairs of users based on their interaction on one social network in order to predict a link on the other. This entails using mobility interactions to predict social links on Twitter, and using social interactions on Twitter to predict links on Foursquare. Subsequently, we are interested in predicting the multiplex links at the cross-section of the two networks using multilayer features. This type of links have both structural and social tie implications as we have demonstrated in this work, which makes them desirable to identify.  

We perform our evaluation on three datasets described at the start of this work in Section 5, where we have Twitter, Foursquare, and the derived multilayer features for the cities of San Francisco, Chicago, and New York. We adopt a supervised learning approach for the prediction tasks, and for each city, which is considered as an independent multilayer network, where we train and test on different layers. Supervised learning methodologies have been proposed as a better alternative to unsupervised models for link prediction \cite{lichtenwalter2010new}.

We compare the performance of feature sets using the Random Forest classifier~\cite{rf2001}  with $10$-fold cross-validation testing strategy: for each test we train on $90\%$ of the data and test on the remaining $10\%$. For every test case the user pairs in the test set were ranked according to the scores returned by the classifiers for the positive class label (i.e., for an existing link), and subsequently, Area Under the Curve (AUC) scores were calculated by averaging the results across all folds. We use AUC scores as a measure of performance because it considers all possible thresholds of probability in terms of true positive (TP) and false positive (FP) values rate, which are computed by comparing the predicted output against the target labels of the test data.

In terms of algorithmic implementation, we have used public versions of the algorithms available in~\cite{pedregosa2011scikit}. 
The features presented earlier in this work, of which each feature set comprises are summarised in Table~\ref{tab:feats}. We denote the Twitter neighbourhood as $\Gamma^T$ and the Foursquare neighbourhood as $\Gamma^F$.
Next, we specify each prediction task and present the results of the supervised learning evaluation in terms of the predictive power of each feature set in both tasks.\\

\begin{table}[h!]
\centering
\begin{tabular}{|r|l|}
\hline
\multicolumn{2}{|c|}{Twitter features}\\
\hline
$mentions$ & $|mentions_{ij}|$ \\ 
\hline
$hashtags$ & $|hashtags_{ij}|$\\
\hline
 $overlap$ & $\frac{|\Gamma_i^T \cap \Gamma_j^T|}{|\Gamma_i^T \cup \Gamma_j^T|}$\\ 
 \hline
 $aa\_sim$&$ \sum\limits_{z \in \Gamma_i^T \cap \Gamma_j^T} \frac{1}{log(|\Gamma_z^T|)}$\\ 
  \hline \hline
\multicolumn{2}{|c|}{Foursquare features}\\
  \hline
$colocs$ & $|colocations_{ij}|$ \\
\hline
$dist$ & $haversine(lat_i,lon_i, lat_j,lon_j)$\\
\hline
$overlap$ & $ \frac{|\Gamma_i^F \cap \Gamma_j^F|}{|\Gamma_i^F \cup \Gamma_j^F|}$\\
\hline
$aa\_sim$ & $\sum\limits_{z \in \Gamma_i^F \cap \Gamma_j^F} \frac{1}{log(|\Gamma_z^F|)}$\\
\hline\hline
\multicolumn{2}{|c|}{Multilayer features}\\
\hline
$int_{GNij} $ &  $\sum\limits_{\alpha}^{M} |int^\alpha_{ij}|$\\
\hline
$sim_{GNij}$ & $\frac{sim_{ij}^a}{dist_{ij}^b}$\\
\hline
$overlap$ & $ \frac{|\Gamma_{CNi} \cap \Gamma_{CNj}|}{|\Gamma_{CNi} \cup \Gamma_{CNj}|}$\\
\hline
$aa\_sim$ & $\sum\limits_{z \in \Gamma_{CNi} \cap \Gamma_{CNj}} \frac{1}{log(|\Gamma_{CNz}|)}$\\
\hline
\end{tabular}
\caption{Summary of link features.}
\label{tab:feats}
\end{table}

\subsection{RQ3: Cross-network prediction}

The Receiver Operating Characteristic (ROC) curves (defined as the True Positive versus False Positive Rate for varying decision thresholds) and the corresponding Area Under the Curve (AUC) scores are shown in Figure~\ref{fig:roc1} for the three datasets. We now discuss these results with respect to each task. In the first prediction task, for a pair of users $i$ and $j$ we define a feature vector $\mathbf{x_{ij}^\alpha}$ encoding the values of the users' feature scores on layer $\alpha$ in the multilayer network. We also specify a target label $y_{ij}^\beta \in \{-1,+1\}$ representing whether the user pair is connected on the $\beta$ layer under prediction.

We use the supervised Random Forest classifier (45 trees, optimised with tree depth = 25) to predict links from one layer using features from the other. Figure~\ref{fig:roc_fsq} shows the ROC curves and respective AUC scores for each dataset in predicting Foursquare links from Twitter features, ranging between 0.7 for the New York dataset to 0.81 for Chicago, and 0.73 for San Francisco. We compare this to the reverse task of predicting Twitter links using Foursquare features in Figure~\ref{fig:roc_mfsq0}, where we obtain AUC scores of 0.86, 0.73, and 0.79 for the three cities respectively. We observe slightly higher results for Twitter links, and we note that this may be as a result of the higher number of Twitter links in our dataset or as a result of the greater difficulty of the inverse task.\\

\subsection{RQ4: Multiplex link prediction}

In our second prediction task, we are interested in evaluating the performance of each feature set in predicting link multiplexity. Given a feature vector $\mathbf{x_{ij}}$, we would like to predict a target label $y_{ij} \in \{-1,+1\}$, where a link exists on both layers (+1)  or none (-1). We compare performance of the multilayer features to the Twitter and Foursquare sets.

In this task, we use all three feature sets to predict multiplex links, which generally exhibit signs of a stronger online bond through interaction and structural properties as we have seen in the first part of this work. In Figures~\ref{fig:m_fsq} and \ref{fig:rm_twt}, we observe how Twitter and Foursquare features perform in predicting multiplex links using the Random Forest algorithm again, with the highest AUC scores of 0.82 and 0.84 for each set respectively. The Foursquare feature set performs better in terms of AUC scores but the multilayer feature set outperforms both (AUC = 0.88 for Chicago), due to its inclusion of features from different layers and cross-layer structural properties. 

In conclusion, it is possible to predict links between heterogeneous social networks and to predict multiplex links spanning multiple networks using multilayer features as we have seen in our subset of users. We discuss the applications of these results in the following section.

\section{Discussion \& Conclusions}

In this work we have demonstrated the structural and interaction properties of links across two online social networks and have also shown the value of multilayer features in predicting links on both Twitter and Foursquare, and multiplex links. We believe that the primary contribution is methodological, since it provides a novel framework for investigating multiplexity across different social networks. The techniques discussed in this work are general and can be potentially used to investigate other scenarios for which datasets containing information about social interactions across multiple networks are available.
In this section, we discuss the implications, limitations and real-world applications of these results.

\subsection{Implications}

Recently, social media has been increasingly alluded to as an \emph{ecosystem}. The parallel comes after the emergence of multiple  OSNs, interacting as a system, while competing for the same resources - users and their attention.  We have addressed this system aspect by modelling multiple social networks as a multilayer online social network in this work. 
We have also identified two extensions of the node neighbourhood. The global neighbourhood or degree gives insight into a users' full connectivity across services, this is especially important when considering users with asymmetric activity and degree across networks since their centrality in the online ecosystem can be under or over-estimated. We additionally defined the core degree, which on the other hand reveals the intersection across networks, and therefore the stronger online ties - those relevant on multiple networks. 

The strength of ties manifested through multiplexity is expressed through a greater intensity of interactions and greater similarity across attributes both the offline~\cite{hay2005,me2014}, and in the online context as we have seen in this work. We have introduced a number of features, which take into consideration the multilayer neighbourhood of users in OSNs. The Adamic/Adar coefficient of neighbourhood similarity in its core neighbourhood version proved to be a strong indicator of multiplex ties. Additionally, we introduced combined features, such as the global interaction and similarity over distance, which reflect more distinctively the type of link, which exists between two users, than its single-layer counterparts. These features can be applied across multiple networks and can be flexible in their construction according to the context of the OSNs under consideration. 



\subsection{Limitations}

Media multiplexity is fascinating from the social networks perspective as it can reveal the strength and nature of a social tie given the full communication profile of people across all media they use~\cite{hay2005}. Unfortunately, full online and offline communication profiles of individuals were not available and our analysis is limited to two social networks. Nevertheless, we have observed some evidence of media multiplexity manifested in the greater intensity and structural overlap of multiplex links and have gained insight into how we can utilise these properties for link prediction. Certainly, considering more OSNs and further relating media multiplexity to its offline manifestation is one of our future goals, and we believe that with the further integration of social media services and availability of data this will be possible in the near future.

Our data is limited to a sub-sample of users who we know have active accounts on both networks in three US cities, Foursquare  check-ins also being limited to those posted on Twitter. This excludes a number of users who may have Foursquare accounts but have not linked them on Twitter. Nevertheless, we were able to show that it is possible to predict one social network from the other in a cross-network manner and we hope to extend our prediction and analysis to a greater scale and geographical scope in the future. \\

\subsection{Applications}

Most new OSNs use contact list integration with external existing networks, such as copying friendships from Facebook through the open graph protocol.\footnote{\url{https://developers.facebook.com/docs/opengraph}} Copying links from pre-existing social networks to new ones results in higher social interaction between copied links than between links created natively in the platform~\cite{Zhong2014}. We propose that extending this copied network with a rank of relevance of contacts using multiplexity can provide even further benefits for newly launched services. 

In addition to fostering multiplexity, however, new OSNs and especially interest-driven ones such as Pinterest for example, may benefit from similarity-based friend recommendations. In this work, we apply mobility features and neighbourhood similarity from Foursquare to predict links on Twitter and vice versa, highlighting the relationship between similar users across heterogeneous platforms. Similarly in~\cite{Tang2012}, the authors infer types of relationships across different domains such as mobile and co-author networks. Although using a transfer knowledge framework, and not exogenous interaction features like we do, the authors also agree that integrating social theory in the prediction framework can greatly improve results. The present work is a step towards understanding the composite nature of online social network services and hopefully towards enhancing their functionality and purpose. 


\section{Acknowledgements}
This work was supported by the Project LASAGNE, Contract No. 318132 (STREP), funded by the European Commission.

\bibliographystyle{aaai}
{\small
\bibliography{biblio}}

\end{document}